\documentclass[iop,revtex4,a4paper]{emulateapj}
\usepackage{amsmath}
\usepackage{lscape}
\usepackage[caption=false]{subfig}
\begin{document}

\title{Evidence of a Bottom-heavy Initial Mass Function in Massive Early-type Galaxies from Near-infrared Metal Lines\footnotemark[*]} \footnotetext[*]{This paper includes data gathered with the 6.5 meter Magellan Telescopes located at Las Campanas Observatory, Chile.}
\shorttitle{Near-IR IMF Tracers}
\shortauthors{Lagattuta et al.}
\author{David J. Lagattuta\altaffilmark{1}}\email{david-james.lagattuta@univ-lyon1.fr}
\author{Jeremy R. Mould\altaffilmark{2,3}}
\author{Duncan A. Forbes\altaffilmark{2}}
\author{Andrew J. Monson\altaffilmark{4,5}}
\author{Nicola Pastorello\altaffilmark{2}}
\author{S. Eric Persson\altaffilmark{4}}
\altaffiltext{1}{Univ Lyon, Univ Lyon1, Ens de Lyon, CNRS, Centre de
  Recherche Astrophysique de Lyon UMR5574, F-69230, Saint-Genis-Laval,
  France}
\altaffiltext{2}{Centre for Astrophysics \& Supercomputing, Swinburne
  University, Hawthorn, VIC 3122, Australia}
\altaffiltext{3}{ARC Centre of Excellence for All-sky Astrophysics
  (CAASTRO)}
\altaffiltext{4}{Observatories of the Carnegie Institution of
  Washington, 813 Santa Barbara Street, Pasadena, CA 91101, USA}
\altaffiltext{5}{Department of Astronomy \& Astrophysics, The
  Pennsylvania State University, 525 Davey Lab, University Park, PA
  16802, USA}


\begin{abstract}
We present new evidence for a variable stellar initial mass function
(IMF) in massive early-type galaxies, using high-resolution,
near-infrared spectroscopy from the Folded-port InfraRed Echellette
spectrograph (FIRE) on the Magellan Baade Telescope at Las Campanas
Observatory.  In this pilot study, we observe several
gravity-sensitive metal lines between 1.1 $\mu$m and 1.3 $\mu$m in
eight highly-luminous ($L \sim 10 L_*$) nearby galaxies.  Thanks to
the broad wavelength coverage of FIRE, we are also able to observe the
\ion{Ca}{2} triplet feature, which helps with our analysis.  After
measuring the equivalent widths (EWs) of these lines, we notice mild
to moderate trends between EW and central velocity dispersion
($\sigma$), with some species (\ion{K}{1},\ion{Na}{1},\ion{Mn}{1})
showing a positive EW-$\sigma$ correlation and others
(\ion{Mg}{1},\ion{Ca}{2},\ion{Fe}{1}) a negative one.  To minimize the
effects of metallicity, we measure the ratio $R$ =
[EW(\ion{K}{1})/EW(\ion{Mg}{1})], finding a significant systematic
increase in this ratio with respect to $\sigma$.  We then probe for
variations in the IMF by comparing the measured line ratios to the
values expected in several IMF models.  Overall, we find that low-mass
galaxies ($\sigma \sim 100$ km s$^{-1}$) favor a Chabrier IMF, while
high-mass galaxies ($\sigma \sim 350$ km s$^{-1}$) are better
described with a steeper (dwarf-rich) IMF slope.  While we note that
our galaxy sample is small and may suffer from selection effects,
these initial results are still promising.  A larger sample of
galaxies will therefore provide an even clearer picture of IMF trends
in this regime.
\end{abstract}

\section{Introduction}
The initial mass function (IMF) of stars is a key component in many
areas of astrophysics, ranging from broad-scope topics such as galaxy
formation and evolution \citep{dav12}, to more specific subjects like
metallicity and star-formation history \citep{mar13}. While many
attempts to construct a ``universal'' IMF describing all galaxies have
been made (see, e.g., \citet{sal55,kro01,cha03}), recent work suggests
that the IMF may actually vary as a function of mass,
\citep[e.g.,][]{tre10,cvd12b,lab13,fer15,spi15} with high-mass
galaxies becoming increasingly ``bottom-heavy'', i.e., described by an
IMF with a larger-than-expected fraction of dwarf stars.
Strengthening this argument is the fact that two largely independent
methods: spectroscopic probes using gravity-sensitive line
measurements and dynamical probes (such as gravitational lensing or
stellar dynamics) measuring the stellar mass-to-light ratio $\Upsilon$
\citep[e.g.,][]{tre10,dut12,cap13,pos15,lei16}, have seen this trend.
If true, this could have a significant effect on scaling relations
such as the Fundamental plane \citep{dut13,mou14}, altering our
understanding of galaxy dynamics and dark matter mass
fractions. Controversy on this subject remains very active, however
(see, e.g., \citealt{smi14,cla15,mar15b,lyu16}).

It is worth noting, though, that spectroscopy-based studies often
limit themselves to optical wavelengths, relying on a relatively small
number of line features that lie below 1 $\mu$m. However,
spectroscopic data taken at near-infrared (NIR) wavelengths have a
number of advantages over their optical counterparts, especially where
IMF-sensitive line features are concerned. In particular, several NIR
absorption lines are gravity-sensitive \citep{mcl07,des12,smi15} and
have significantly different line strengths in giant and dwarf stars,
making them well-suited to probe different stellar populations and
expanding the pool of IMF tracers. Similarly, the flux of dwarf stars
peak in the 1 -- 2 $\mu$m range \citep{fro75}, increasing the relative
contribution of dwarf stars to the integrated galaxy light at these
wavelengths. This in turn makes it easier to measure dwarf-sensitive
absorption lines, allowing us to look for even fainter
features. Furthermore, galaxy light is less sensitive to dust at NIR
wavelengths, diminishing the effects of extinction, which can lower
continuum- flux levels and decrease the signal-to-noise ratio (SNR) of
detected lines.

In this paper, we present the initial results of the search for
IMF-sensitive features at NIR wavelengths, using high-resolution
spectroscopic data taken with the Magellan telescope. For our initial
pilot data sample we select eight massive, highly luminous ($L_{\rm
  gal}$ $\sim$ 10 $\times$ L$_{*}$), nearby galaxies, looking for
trends in line strengths and line ratios as a function of galaxy mass.
To fully explore the galaxy mass phase-space, we consider two possible
mass tracers: total K-band luminosity (a proxy for stellar mass) and
central velocity dispersion, a direct probe of the galaxy's potential
well.  Comparing our EW measurements to the predicted EW values of
(IMF-dependent) theoretical models (e.g., \citealt{con10a}) we are
able to account for the effects of IMF-variation on EW, allowing us to
probe IMF-sensitive trends as a function of mass.  Furthermore, by
studying EW ratios in addition to simple line strengths, we are also
able to control for variables such as stellar age and metallicity.

This paper is organized as follows: in Section 2 we present the data
and describe the steps we take to create reduced 1-D spectra.  In
section three we identify the lines used in this study and layout the
procedure for measuring EWs.  We measure the actual EW values and
search more mass-dependent trends in Section 4, and we compare the
results to theoretical models in Section 5.  We briefly summarize our
results and conclude in Section 6.  Throughout this work, we assume a
standard cosmological model with $\Omega_{\Lambda} = 0.7$, $\Omega_M =
0.3$, and Hubble parameter $h = 0.7$

\section{Data and Reduction} 
\subsection{Data acquisition}
We observed our galaxy sample using the Folded-port InfraRed
Echellette spectrograph (FIRE; \citealt{sim08}) on the Magellan Baade
telescope at Las Campanas Observatory.  Data were acquired in
good/fair conditions over several nights in 2014.  The full list of
observations can be seen in Table \ref{tbl:ObsPars}.  Additionally, in
Table \ref{tbl:LitPars} we present the physical parameters of each
galaxy (taken from publicly available data archives), which were used
to motivate the sample selection.  These parameters are: redshift,
central velocity dispersion ($\sigma_c$), total luminosity ($L_{\rm
  NIR}$), angular size ($R_{\rm Tot}$), and half-light radius
($R_{e}$).  For all observations, we used FIRE in echellette mode with
a 0.6\arcsec slit width, providing wavelength coverage from 0.82$\mu$m
to 2.51 $\mu$m at a spectral resolution of R = 6000.  The nominal slit
length of FIRE is 7\arcsec.  In order to maximize sensitivity while
minimizing readout noise, we used the ``High gain'' (1.3 e$^-$
DN$^{-1}$) detector setting, limiting the exposure time of any single
observation to 30 minutes to avoid detector saturation and non-linear
effects.  ``Science'' frames (galaxies and telluric correction stars)
were readout in Sample-Up-The-Ramp mode, while ``Calibration'' frames
(arcs, flats, and twilight sky flats) were readout in single-sample
Fowler mode.

 We observed telluric standards immediately following each target
 galaxy, in order to match atmospheric conditions of science objects
 as closely as possible.  To choose the best available telluric stars
 (which also served as a flux-calibrators), we used
 \texttt{find\_tellurics}\footnote{http://web.mit.edu/$\sim$rsimcoe/www/FIRE/ob\_manual.htm\#Tellurics},
 an internally-provided script that scans a database of known A0V
 stars and looks for matches based on airmass and sky angle.  For our
 galaxy sample, the typical best-matched stars had average V-band
 magnitudes m$_{\rm V}$ $\sim$ 9.5, which was considerably brighter
 than the galaxies themselves.  Thus, to again avoid saturation
 effects we only observed each telluric star for a total of 60 seconds
 (integrated over two 30-second exposures), along with a Thorium-Argon
 arc spectrum to provide a wavelength solution.

\begin{deluxetable*}{ccccccc}
\tablecaption{Observing Log. \label{tbl:ObsPars}}
\tablehead{  
\colhead{Target Name} & \colhead{RA} & \colhead{Dec} & \colhead{Obs. Date} & \colhead{Exp. Time (s)} & \colhead{Mean Airmass} & \colhead{Mean Seeing (\arcsec)}}

\startdata
NGC 1316 & 03:22:41.789 & -37:12:29.52 & 2014 12 12 & 5400 & 1.24 & 0.53 \\
NGC 1332 & 03:26:17.321 & -21:20:07.33 & 2014 12 13 & 5400 & 1.32 & 0.79 \\
NGC 3258 & 10:28:53.588 & -35:36:19.98 & 2014 12 13 & 3600 & 1.16 & 0.75 \\
NGC 3557 & 11:09:57.653 & -37:32:21.02 & 2014 12 12 & 3600 & 1.28 & 0.42 \\
NGC 5845 & 15:06:00.787 & +01:38:01.77 & 2014 06 15 & 3600 & 1.22 & 0.93 \\
NGC 7014 & 21:07:52.185 & -47:10:44.53 & 2014 06 15 & 7200 & 1.23 & 0.96 \\
NGC 7410 & 22:55:00.945 & -39:39:40.93 & 2014 09 14 & 5400 & 1.20 & 0.80 \\
NGC 7743 & 23:44:21.130 & +09:56:02.55 & 2014 06 15 & 7200 & 1.45 & 1.55 
\end{deluxetable*}

\begin{deluxetable*}{ccccccc}
\tablecaption{Archival Galaxy Parameters. \label{tbl:LitPars}}
\tablehead{
  \colhead{Galaxy} &  \colhead{$z$}\tablenotemark{a} & \colhead{$\sigma_c$}\tablenotemark{b} & \colhead{$L_{\rm NIR}$}\tablenotemark{c} & \colhead{$R_{\rm Tot}$}\tablenotemark{c} & \colhead{$R_{e}$}\tablenotemark{c} & \colhead{Slit Coverage}\tablenotemark{d}\\
\colhead{} & \colhead{} & \colhead{(km s$^{-1}$)} & \colhead{($10^{12} L_{\odot}$)} & \colhead{(\arcsec)} & \colhead{(\arcsec)} & \colhead{(\%)}}

\tablenotetext{a}{Average value from the NASA/IPAC Extragalactic Database (NED)}
\tablenotetext{b}{Taken from the HYPERLEDA archive}
\tablenotetext{c}{Averaged from J-, H-, and K-band data in the 2MASS eXtended Source Catalog (XCS)}
\tablenotetext{d}{Fraction of Effective radius ($R_{e}$) covered by the FIRE slit}
\startdata
NGC 1316  & 0.00588  & 224.5  & 2.591  & 295.00  & 46.62  &  7.5\\
          & (0.00005)  & (3.3)  & (0.029)  &     &  (2.29)  &   \\
NGC 1332  & 0.00510  & 312.5  & 0.504  & 182.50  & 27.02  & 13.0\\
          & (0.00004)  & (10.7)  & (0.006)  &    &  (0.66)  &   \\
NGC 3258  & 0.00925  & 257.0  & 0.526  &  73.59  & 15.81  & 22.1\\
          & (0.00005)  & (12.7)  & (0.007)  &    &  (0.58)  &   \\
NGC 3557  & 0.01023  & 260.1  & 2.546  & 126.37  & 26.13  & 13.4\\
          & (0.00005)  & (7.9)  & (0.025)  &     &  (0.77)  &   \\
NGC 5845  & 0.00490  & 230.2  & 0.070  &  40.39  &  4.78  & 73.2\\
          & (0.00006)  & (8.1)  & (0.001)  &     &  (0.09)  &   \\
NGC 7014  & 0.01618  & 292.8  & 0.741  &  56.46  & 11.21  & 31.2\\
          & (0.00004)  & (7.7)  & (0.011)  &     &  (0.42)  &   \\
NGC 7410  & 0.00580  & 134.0  & 0.561  & 184.40  & 47.67  &  7.3\\
          & (0.00009)  & (12.4)  & (0.006)  &    &  (0.43)  &   \\
NGC 7743  & 0.00562  &  84.7  & 0.176  &  93.23  & 29.97  & 11.7\\
          & (0.00004)  & (2.3)  & (0.003)  &     &  (0.87)  &     
\end{deluxetable*}

\subsection{Reduction}
Data reduction is primarily handled by FIREHOSE, an IDL-based pipeline
designed specifically for FIRE data.  While much of the pipeline is
automated, there are a few times where we deviated from the standard
output, which we briefly describe.

The pipeline first creates ``master'' quartz-lamp pixel-flats (for
flat-fielding) and twilight flats (for illumination correction) from
initial calibration data.  During each observing night, we took five
pixel-flat frames in the afternoon calibration session and two
sky-flats during dawn twilight.  The one exception to this was the
first observing night (2014 June 05) where we took no sky-flat data,
forcing us to substitute other sky frames for the reduction.  After
testing several options (either using sky frames taken during the
other observing nights or publicly available archival data) we found
no significant differences in any of the generated flat-field frames
or science outputs.  Therefore, in the final reduction for that
night's data we used the flat field files from the second observing
night (2014 September 09), as it was the closest date to the original
run.

After flat fielding, the pipeline extracts the spectral trace from
each frame, using an iterative process to fit the object profile,
subtract sky flux, and attach a wavelength solution.  We model the sky
flux from the science frames themselves instead of separate sky
frames, using the outermost 1\arcsec\ regions at either end of the
slit.  This is to better account for the highly time-variable nature
of the NIR sky.  The extraction method is user-selected, and can take
the form of either an optimal, non-parametric window function or a
simple boxcar function.  We experimented with both methods for our
data, comparing the results of each case.  The optimal extraction
method produced spectra that were less noisy and contained fewer
cosmic ray artifacts, but (even after flux calibration) had very
disjoint inter-order continuum flux levels.  Conversely, the boxcar
method had much smoother continuum levels, but were noisier.  To
obtain the best reduction possible, then, we combine elements of each
extraction method in the final product.  In particular, we use the
optimally-extracted traces, but re-scale them to the boxcar-extracted
continuum levels before analyzing them.

\begin{figure}
\includegraphics[width=8.8cm]{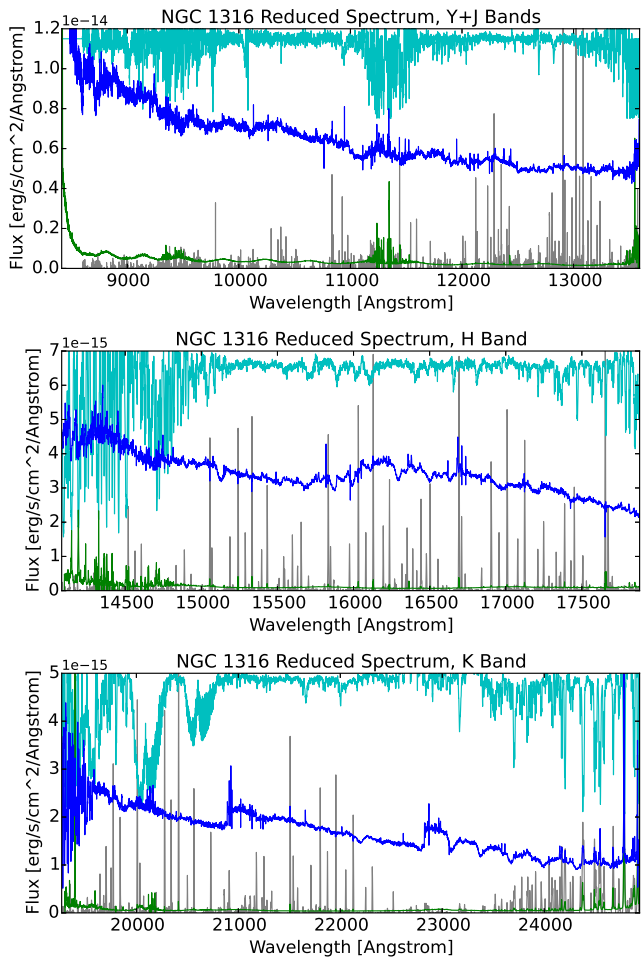}
\caption{Reduced 1D spectrum of the nuclear region of NGC 1316, the
  massive AGN-host galaxy in the Fornax cluster, and one of the eight
  galaxies used in this study.  The observed spectrum is shown in
  blue, while the error spectrum (multiplied by a factor of 10) is
  shown in green.  The error spectrum is upscaled simply to make it
  more visible in the plot.  We also show a scaled sky emission
    spectrum (gray) and telluric absorption spectrum (cyan) to better
    highlight areas of increased sky noise.  In the ``cleanest''
    wavelength regions with few sky lines and high atmospheric
    transmission, the typical SNR/Angstrom is $\sim$ 100.  In other
    regions where sky noise is high or instrument sensitivity is low
    (at bluer wavelengths) the SNR can fall to as low as 10 --
    20. \label{fig:Fspec}}
\end{figure}

Following extraction, the data are flux-calibrated and corrected for
telluric absorption using the procedure implemented in the xtellcor
package (Vacca, Cushing,\& Rayner 2003).  Finally, the extracted,
calibrated 2D data are collapsed to 1D, and all data frames for
a given science target are combined using a weighted average.
Individual echelle orders are then stitched together (using an
inverse-variance weighting scheme) into a final spectrum.  An example
of the final output can be seen in Figure \ref{fig:Fspec}.

\section{Line Measurements}
\subsection{Line Selection}
\subsubsection{M-dwarf Metal Lines}
To probe for variations in the IMF, we look for M-dwarf absorption
features in our galaxy spectra.  \citet{mcl07} and \citet{des12}
identify a number of these features between 1.1 and 1.3 micron, giving
us a wide range of lines to choose from.  In order to avoid the
complicated physics of molecular rotation, however, we limit ourselves
to neutral atomic lines.  Following \citet{des12}, we only select
lines with an oscillator strength $\log$(gf) $>$ -2 and continuum
line-depth D $>$ 0.2, as this maximizes the chance that the feature
will be detected in the integrated galaxy spectrum. After applying
these criteria, we are left with 18 candidate absorption lines.  The
full list of features, along with rest-frame wavelengths (measured in
air) is shown in Table \ref{tbl:DLines}.

To check for IMF-variability in these features, we generate a series
of synthetic stellar population (SSP) models with different IMF slopes
and compare line strengths between models. To remove metallicity
effects to first order, we sum all lines of a given element together
and then divide by the combined \ion{Fe}{1} line strength.  This makes
the results largely insensitive to metallicity in the
--0.3$<$[Fe/H]$<$0 range of models.  We do note that additional
testing shows that the strength of these lines is much more dependent
on IMF and stellar age than metallicity (see section \ref{Models}),
suggesting that this normalization procedure will not significantly
change the results.  However, we choose to include it in this work
anyway, to remove as many external factors from the analysis as
possible.  From this exercise, we find that \ion{K}{1} shows the most
IMF sensitivity, though other lines (such as \ion{Mg}{1}, \ion{Na}{1},
and \ion{Al}{1}) also show moderate trends.  As expected, these are
low ionization potential elements.  Because of this, much of the
analysis and discussion in this work focuses on these select features
(especially \ion{K}{1}), though we include the results of all the
candidate lines for completeness.

\begin{deluxetable}{rcl}
\tablecaption{Target M-dwarf Atomic Line Features \label{tbl:DLines}}
\tablehead{  
  \colhead{Element} & \colhead{Rest-Frame Wavelength}\tablenotemark{a} & Plot Color\tablenotemark{b}\\
  \colhead{} & \colhead{(Angstrom)}
}
\tablenotetext{a}{Measured in air}
\tablenotetext{b}{Color used to identify this element group in subsequent plots}
\startdata
\ion{K}{1}   &  11690.2  &  black\\
\ion{K}{1}   &  11769.6  &  black\\
\ion{K}{1}   &  11772.8  &  black\\
\ion{Fe}{1}  &  11783.3  &  magenta\\
\ion{Mg}{1}  &  11828.1  &  cyan\\
\ion{Fe}{1}  &  11882.8  &  magenta\\
\ion{Ti}{1}  &  11892.9  &  orange\\
\ion{Fe}{1}  &  11973.0  &  magenta\\
\ion{K}{1}   &  12432.3  &  black\\
\ion{K}{1}   &  12522.1  &  black\\
\ion{Na}{1}  &  12679.1  &  red\\
\ion{Ti}{1}  &  12738.4  &  orange\\
Pa-$\beta$   &  12821.7  &  green\\
\ion{Ti}{1}  &  12847.0  &  orange\\
\ion{Mn}{1}  &  12899.8  &  brown\\
\ion{Al}{1}  &  13123.4  &  gray\\
\ion{Ca}{1}  &  13134.9  &  yellow\\
\ion{Al}{1}  &  13150.8  &  gray   
\end{deluxetable}

\subsubsection{The Calcium-II Triplet}
Thanks to the large wavelength range of FIRE, we can also detect the
\ion{Ca}{2} triplet feature at $\lambda$[8498,8542,8662] $\AA$, which
provides a number of added benefits to the analysis.  First,
\ion{Ca}{2} is known to be sensitive to the IMF \citep{sag02,cen03}, giving
us an independent check to the M-dwarf analysis.  Second, the
line-depth of \ion{Ca}{2} is significantly stronger than the dwarf
lines, making identification much easier.  Finally, the SNR of the
feature (again considerably higher than the dwarf lines) is high
enough to estimate a velocity dispersion, which can be used as a proxy
for mass.

\begin{figure*}
  \centerline{
    \includegraphics[width=8.8cm]{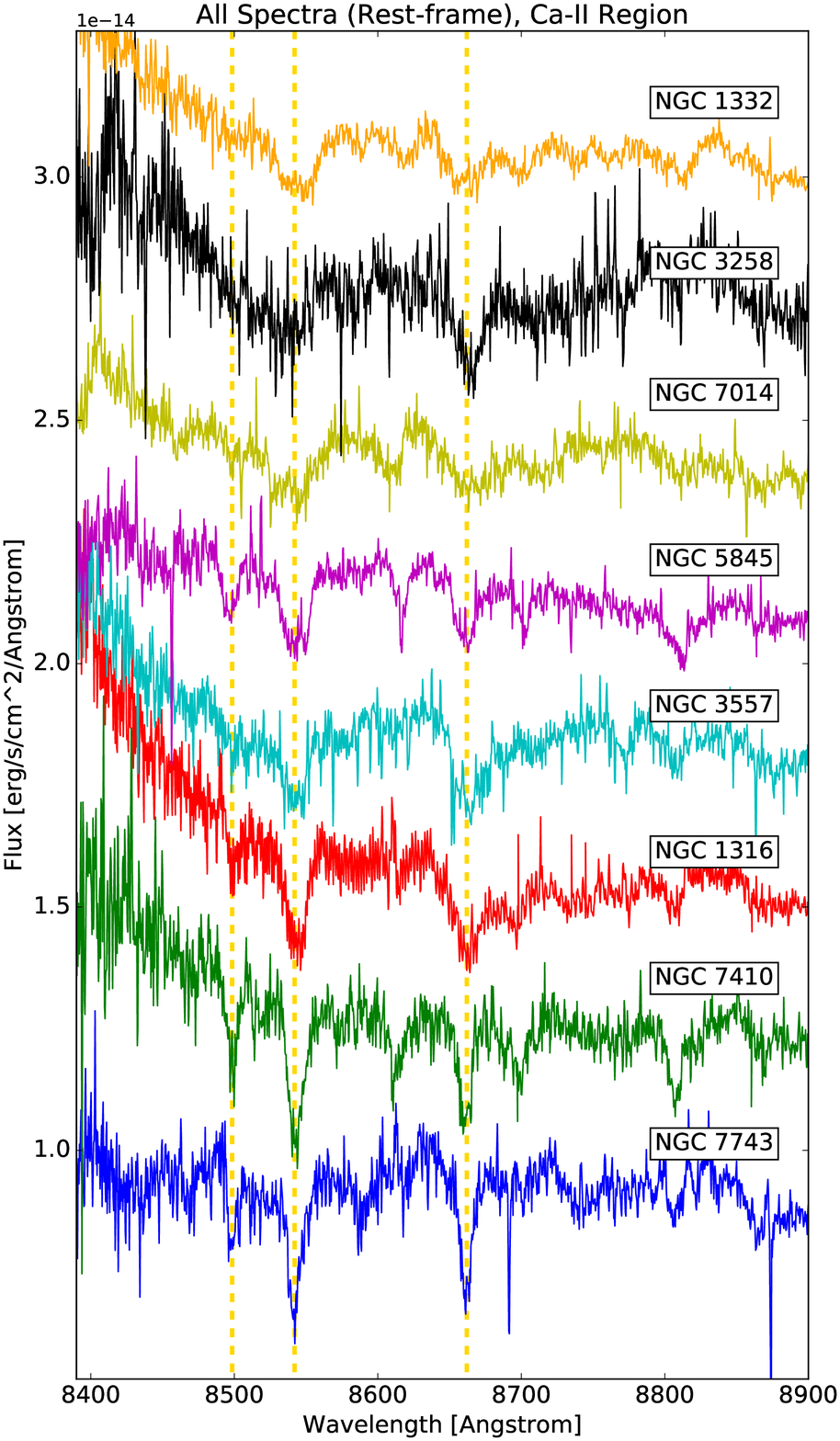}
    \includegraphics[width=8.8cm]{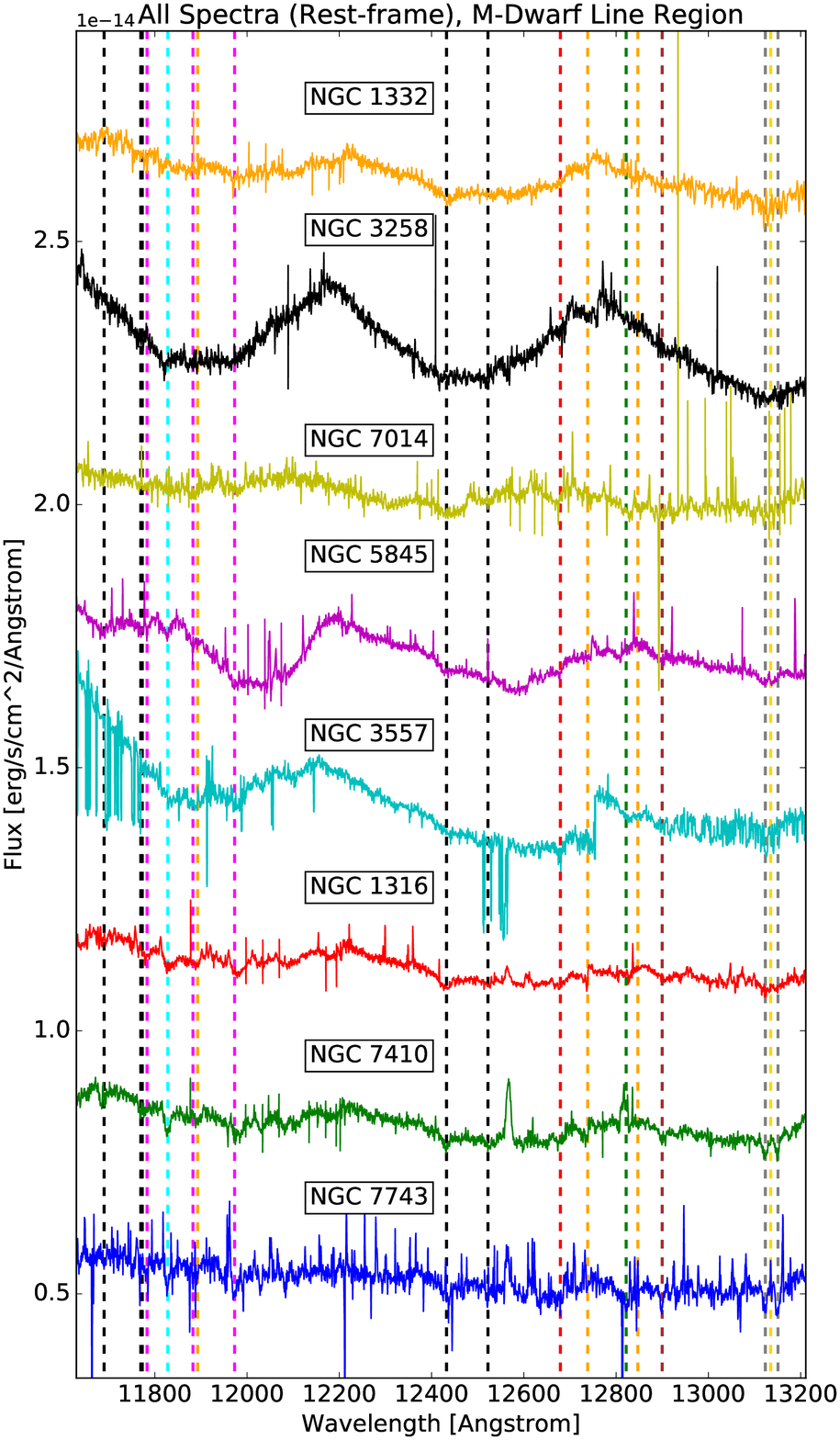}}
  \centerline{
    \includegraphics[width=17.6cm]{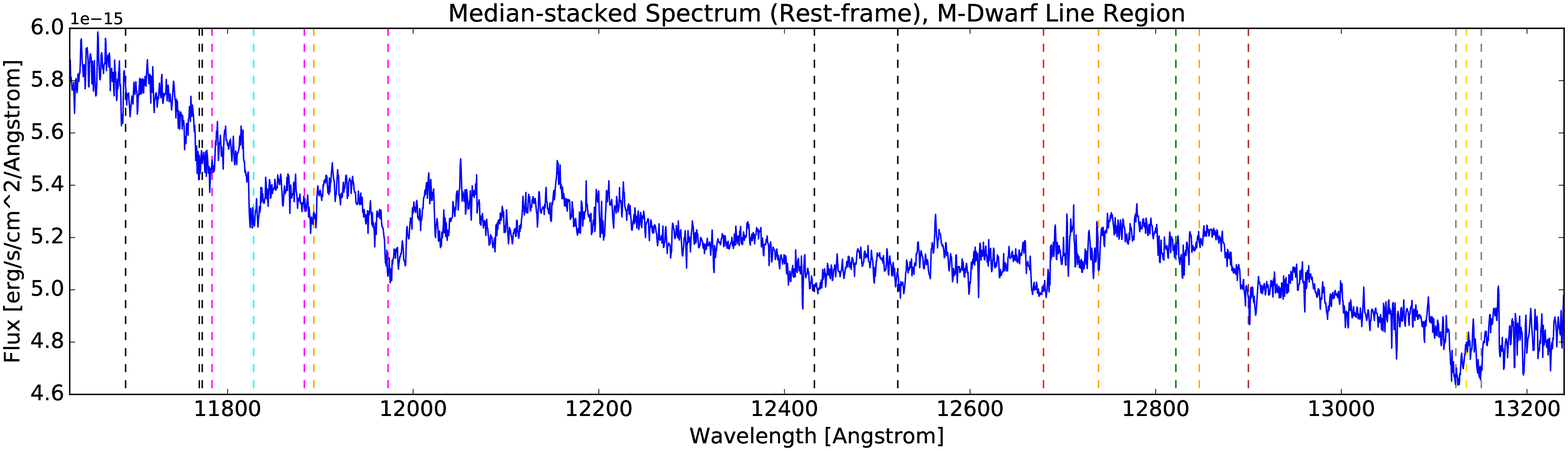}}
  \caption{Cutout spectral regions from each galaxy in our sample,
    identifying the lines used to probe for IMF variations.  {\bf Top
      Left:} The portion of the spectra containing the strong
    \ion{Ca}{2}-triplet.  Yellow dashed lines identify each absorption
    feature.  {\bf Top Right:} The spectral region containing the
    faint M-dwarf lines.  Dashed lines again represent the location of
    absorption features, and follow the order of elements presented in
    Table \ref{tbl:DLines}.  The element identified by each dashed
    line is given by its color, according to the following: Black =
    \ion{K}{1}, Magenta = \ion{Fe}{1}, Cyan = \ion{Mg}{1}, Orange =
    \ion{Ti}{1}, Red = \ion{Na}{1}, Green = Pa$\beta$, Brown =
    \ion{Mn}{1}, Grey = \ion{Al}{1}, Yellow = \ion{Ca}{1}.  Spectra
    are ordered from bottom to top based on increasing central
    velocity dispersion, and a distinct trend between dispersion and
    line-depth (especially with \ion{Ca}{2}) is clearly visible.  To
    make line identification easier, we shift each spectrum to the
    rest-frame.  Additionally, each spectrum has been arbitrarily
    offset along the flux axis to prevent features from
    overlapping. {\bf Bottom:} The median-stacked average of all
    spectra, covering the M-dwarf line region.  Here, the
    absorption features used in this work are clearly
    visible. \label{fig:allLines}}
\end{figure*}

All lines (identified in every galaxy) can be seen in Figure
\ref{fig:allLines}.

\subsubsection{Rejected IMF-Sensitive Tracers}
In addition to the M-dwarf metal lines and \ion{Ca}{2} triplet,
previous studies have identified other IMF-sensitive spectral lines at
NIR wavelengths which could, in principle, be used as additional
constraints.  However, due to redshift, SNR, and/or modeling effects,
were were unable to include them in our data set.

\begin{itemize}

\item The \ion{Na}{1} doublet $\lambda$[8183,8195] (e.g.,
  \citealt{sch97a,van12}) falls slightly outside of the FIRE
  wavelength coverage for our galaxy sample, except for the highest
  redshift galaxies ($z > 0.0007$.)  Since we were not able to measure
  the line for all of our targets, we did not include it in this work.

\item The Wing-Ford $\lambda$9916 FeH band
  (e.g. \citealt{sch97b,mcc16}) is covered by the FIRE spectra, but
  its broad, shallow shape makes it difficult to distinguish from
  local continuum variations.  Additionally, since FeH is a molecular
  feature, we reject it in favor of simpler atomic absorption lines.
  This is because, although the diatomic molecule FeH is included in
  the synthetic spectra of \citet{all12} that we use to compare to
  observation, there is a danger that its detailed line list is
  incomplete. This requires further investigation.  For now we would
  recommend using empirical spectral libraries (such as MILES) where
  FeH is used as an IMF indicator, rather than synthetic. But it is
  also important to study the metallicity dependence of FeH bands, as
  it is likely to follow the behavior of CaH, which is stronger in
  metal poor subdwarfs than solar metallicity main sequence stars.

\item Finally, the \ion{Na}{1} $\lambda$11400 \citep{smi14} line falls
  in a telluric water absorption band (11100 -- 11500 angstrom) for
  all of our galaxy sample redshifts, making it impossible to
  accurately measure its line strength.
\end{itemize}

\subsection{Measuring lines}
\label{Measurements}
\subsubsection{The Ca-II Procedure}
After identifying lines of interest, we then model and measure their
equivalent widths (EWs).  Since the \ion{Ca}{2} lines are always the
strongest absorption features seen, we independently model these
features before moving on the fainter M-dwarf lines.

First, we model galaxy kinematics using the penalized pixel-fitting
routine (pPXF) described in \citet{cap04}.  For reference spectra, we
use the \citet{tre04} galactic templates (constructed from the
\citet{bru03} stellar population libraries), rescaled to match the
resolution of our FIRE data.  This provides an estimate of redshift
and central velocity dispersion.  At the median redshift of our galaxy
sample, the FIRE slit aperture (7\arcsec$\times$0.6\arcsec)
corresponds to a physical size of $\sim$ (1$\times$0.1) kpc.

To make an initial guess for the fit parameters, we use the ``best''
redshift estimate taken from the NASA/IPAC Extragalactic Database
(NED), and small-aperture velocity dispersion measurements from the
HYPERLEDA\footnote{http://leda.univ-lyon1.fr/} archive \citep{mak14}.
In most cases, the best-fit pPXF redshift is within 50 km/s of the NED
value, but there are a few instances (NGC 3258 and NGC 5845) where the
shift is more substantial ($\sim$ 150 km s$^{-1}$).  To test for
asymmetry in the line profile, we also run the pPXF fit with a more
general model, modifying the initial Gaussian fit-shape with
higher-order Gauss-Hermite polynomial terms, following, e.g.,
\citet{van93} and \citet{cap04}.  By including the first two terms
($h_3$ and $h_4$) we do alter the best-fit velocity dispersion, but
this change is small (typically between 5 and 10 km s$^{-1}$) and
always within the measurement uncertainty of the Gaussian fit.  The
redshift of the fit remains virtually unchanged.  Since this result is
minor, we adopt the Gaussian model parameters for the rest of the
fitting procedure, especially since the M-dwarf lines do not have a
high enough SNR to accurately measure profile asymmetry.

After measuring the redshift, we convert the observed spectra to
rest-frame and extract a small (500 $\AA$ ) cutout, centered between
the \ion{Ca}{2} $\lambda$8542 and \ion{Ca}{2}$\lambda$8662 lines.
This provides enough data to model both the \ion{Ca}{2} absorption and
the surrounding continuum flux.  We fit the continuum with a 3rd-order
polynomial spline model, then divide the cutout by this model to
normalize the flux.  After normalizing, we ``invert'' the spectrum
(Inverse = 1 - Normalized) to remove the continuum, leaving only the
inverted absorption features.

The model of the \ion{Ca}{2} system is relatively simple, consisting
of a set of three Gaussians, one for each absorption feature.  While
the amplitudes of each line vary independently, we fix the $\sigma$
value to match the pPXF fit.  Prior to measuring final EWs, however,
we convolve the data and models to a common velocity dispersion
$\sigma = 350$ km s$^{-1}$, in order to account for kinematic
broadening effects.  Additionally, we allow the redshift of the system
to vary, ensuring that the Gaussian peak finds the center of the
absorption features and eliminating any systematic errors in the
wavelength solution.  In general, the difference between the
Gaussian-fit redshift and the pPXF value is small (again $<$ 10 km
s$^{-1}$), but accounting for this extra shift makes later
identification of Dwarf lines much easier, since many of these
features are just at or above the level of the noise leading to
possible misidentification or confusion.

\begin{figure}
  \begin{center}
    \centerline{
      \includegraphics[width=8.8cm]{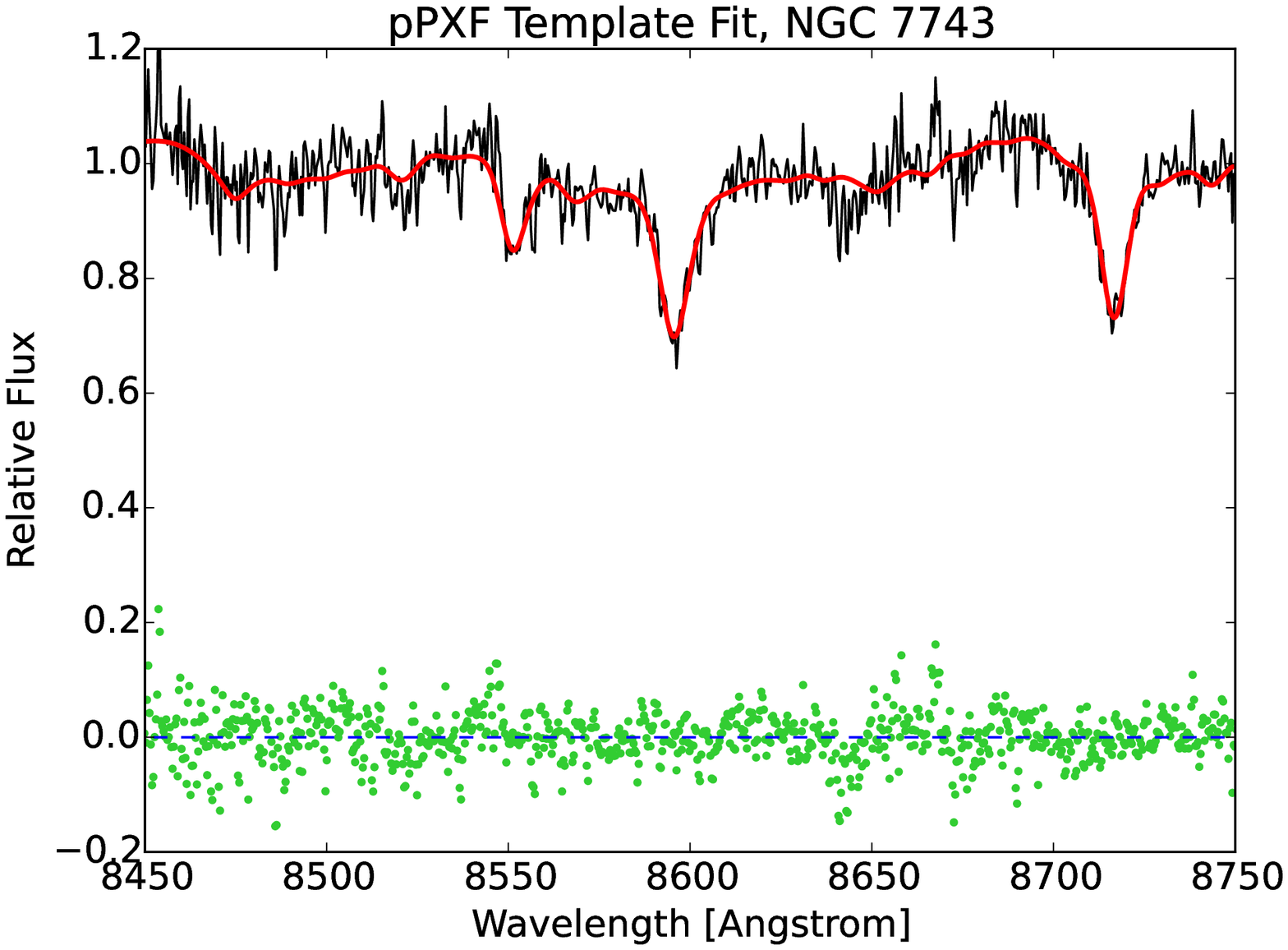}}\hfill~
    \centerline{
      \includegraphics[width=8.8cm]{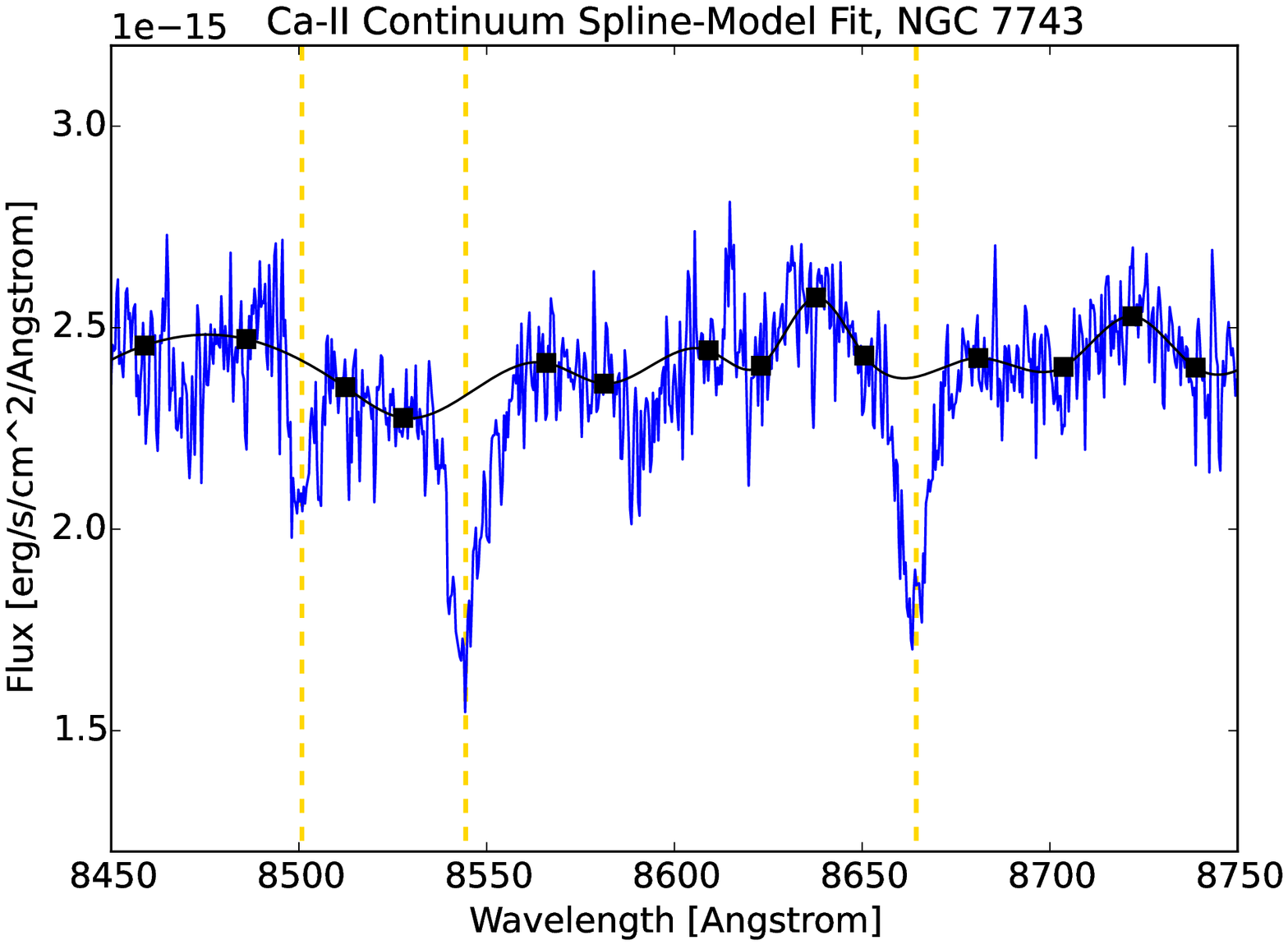}}\hfill~
    \centerline{
      \includegraphics[width=8.8cm]{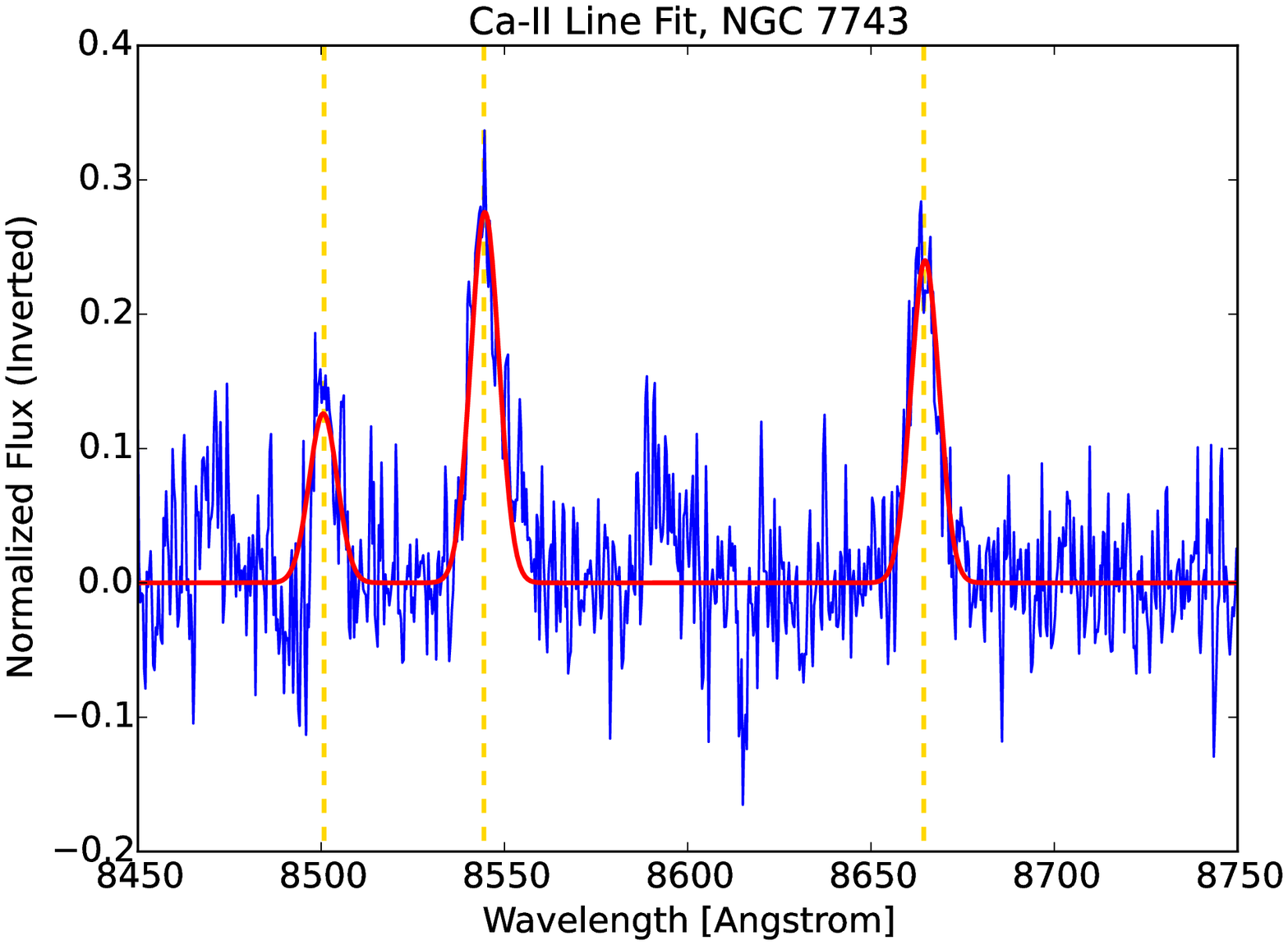}}
  \end{center}
  \caption{Example of the \ion{Ca}{2} fitting procedure, applied to
    NGC 7743.  {\bf Top:} Kinematic and redshift data are fit with the
    pPXF modeling code \citep{cap04}, using \citet{bru03} galaxy
    templates. The real spectrum is shown in black, while the model
    template is shown in red.  Residuals are represented as green
    dots.  {\bf Middle:} We model the continuum flux surrounding
    \ion{Ca}{2}, using a 3rd-order polynomial spline function.  The
    absorption features (represented by dashed yellow lines) are
    clearly visible. Before fitting, the spectrum is shifted to the
    rest-frame using the the best redshift value from the pPXF fit.
    {\bf Bottom:} The spectrum is normalized by the continuum model,
    then inverted (taking 1 - Normalized flux), leaving only the
    \ion{Ca}{2} feature.  We fit this feature with a series of
    Gaussians (using the pPXF $\sigma$ value as the FWHM), which we
    use to measure an equivalent width (EW).  Simultaneously, we
    measure a new best-fit redshift value which we use to better
    identify central positions of M-dwarf absorption lines.  EW values
    obtained from the model fits are shown in Table
    \ref{tbl:modParams}.  \label{fig:CaTFit} }
\end{figure}

The model is optimized though a standard chi-square minimization
process, using a Levenberg-Marquardt algorithm to explore parameter
space.  Once the \ion{Ca}{2} parameters are optimized, we create the
new best-fit \ion{Ca}{2} model and simply sum the flux of each
individual Gaussian to obtain the EW of the line.  To estimate model
uncertainty, we add Poisson noise to the spectrum and re-run the fit.
After repeating this process 500 times, we combine the results, taking
the standard deviation of the set to be the model error.  An example
of the full fitting procedure can be seen in Figure \ref{fig:CaTFit},
and the \ion{Ca}{2} EW values are shown in Table \ref{tbl:modParams}.

To check the accuracy of our measurement technique, we also calculate
EW values for each \ion{Ca}{2} feature using the line-strength indices
presented in Table 1 of \citet{cvd12a}.  We find excellent agreement
between the two methods, and in all cases the difference between the
EW of the line index and our fitting technique is within the
statistical uncertainty of the model.  In some ways this is expected,
since our model-fitting technique is designed to mimic the classical
approach, but with a slightly modified continuum fit.  Regardless, the
agreement suggests our procedure is robust, and that we can
confidently use it to fit the other metal lines, which do not have
formally defined line indices.

\subsubsection{Other lines}
The fitting and measuring procedure for the dwarf lines largely
follows that of \ion{Ca}{2}, though there are some modifications based
on the \ion{Ca}{2} fit.  Starting again with the observed-frame
spectrum, we shift the data to the new rest-frame, as determined by
the best-fit model redshift.  This new redshift accounts for the small
shift between the Gaussian and pPXF fit values, as mentioned in the
previous section.  We then extract the full spectral range containing
the dwarf lines, from $\lambda = 11500 \AA$ to $\lambda = 13500 \AA$
and fit a spline model to measure the continuum.  Rather than fit the
continuum around each line individually, we construct a full spline
model over the entire wavelength range, since many of the lines fall
close to one another, leaving insufficient continuum for a proper fit.
To ensure a consistently ``clean'' set of continuum points for each
galaxy, we make use of the \citet{men15} galaxy templates, looking for
points close to the lines that do not fall on any other absorption or
emission features.  To better match the models to our galaxy sample,
we convolve the initial template with a Gaussian kernel equal to the
galaxy's velocity dispersion (typically around 300 km s$^{-1}$),
making sure that the selected continuum points do not overlap real
signal.  In most cases, the model-selected continuum points are
unchanged from galaxy to galaxy, however in the case of known AGN (NGC
1316, NGC 7410, and NGC 7743) [\ion{Fe}{2}] emission coincides
directly with the continuum point at $\lambda = 12566 \AA$, leading to
a poor fit and affecting the EW measurements of nearby \ion{K}{1}
$\lambda$12522 and \ion{Na}{1} $\lambda$12679.  To avoid this problem,
we simply remove that point from the fit, and replace it with two new
points immediately longward and shortward of the emission line.  This
eliminates the fitting error in the region without significantly
changing the fit in any other location.

\begin{figure}
  \begin{center}
    \centerline{
      \includegraphics[width=8.8cm]{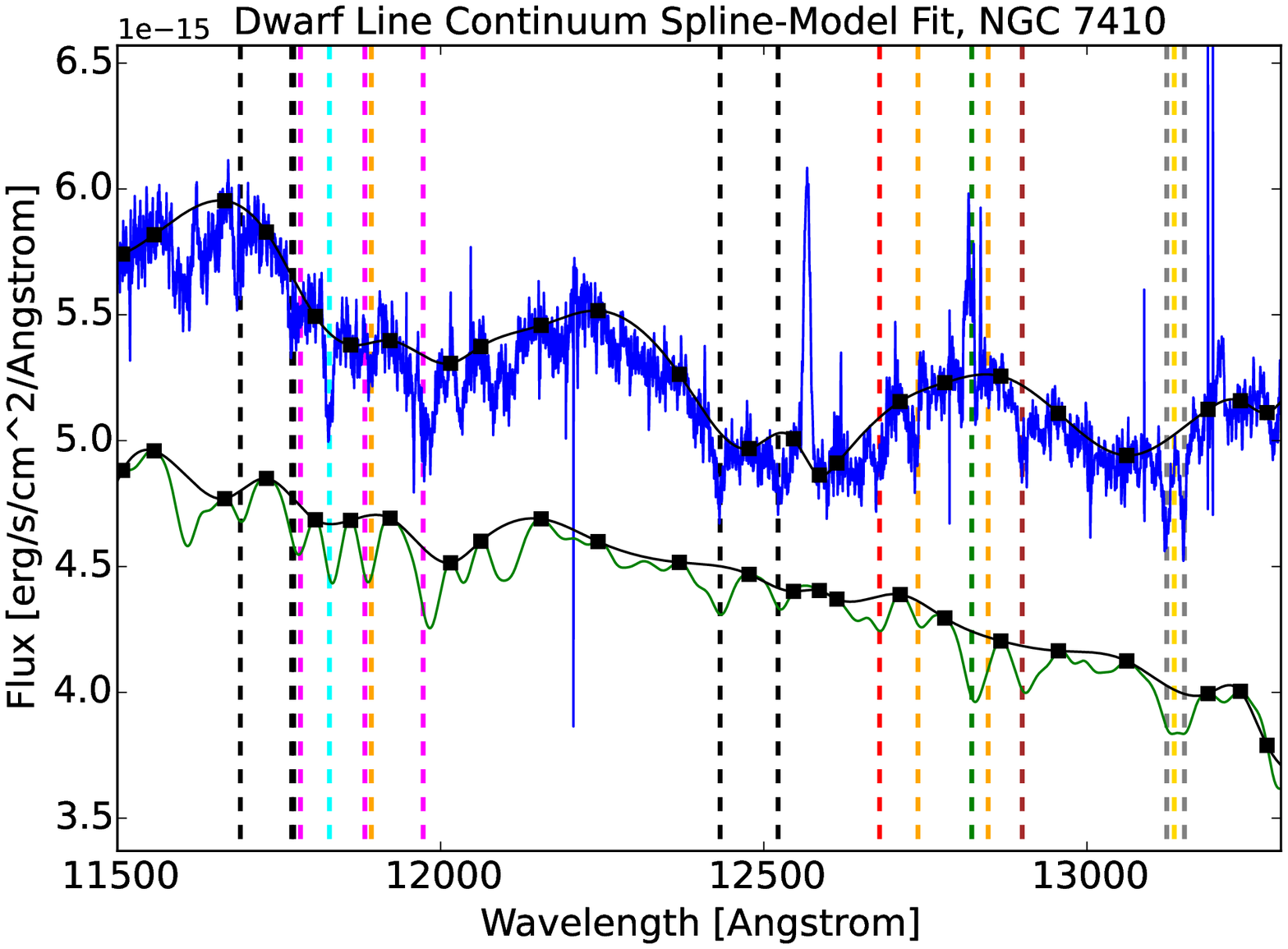}}\hfill~
    \centerline{
      \includegraphics[width=8.8cm]{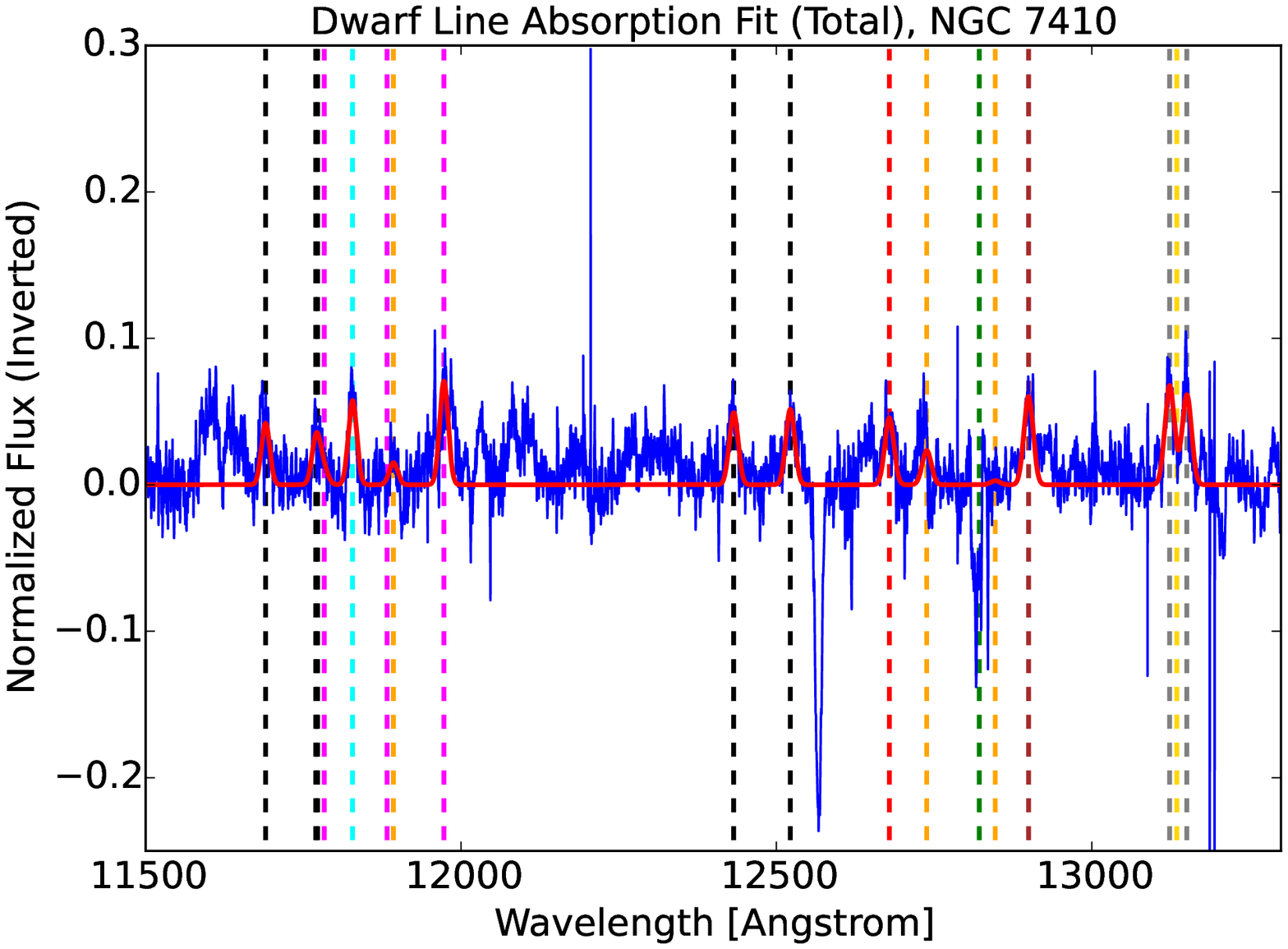}}\hfill~
    \centerline{
      \includegraphics[width=8.8cm]{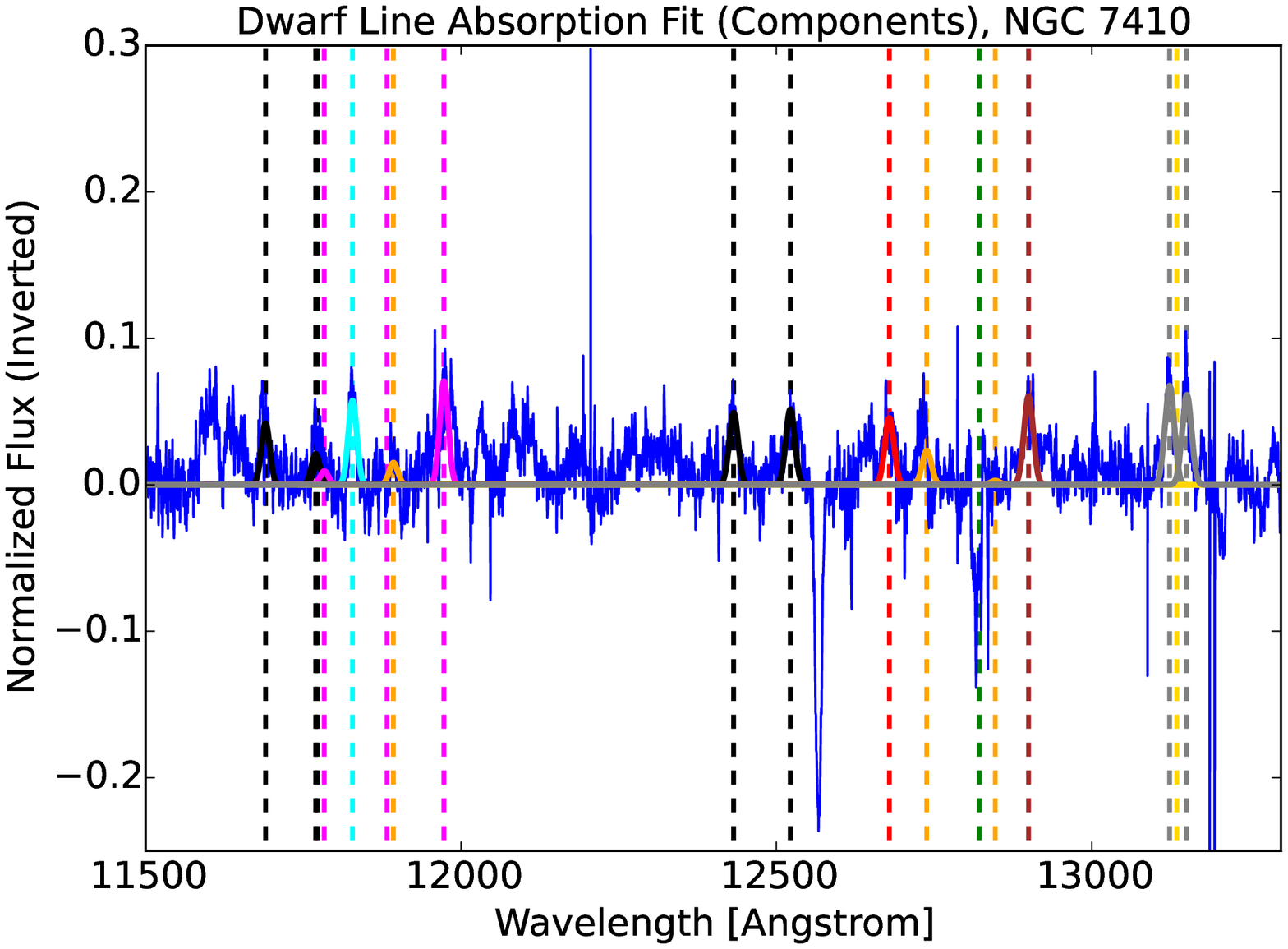}}
    \caption{Example of the M-dwarf line fitting procedure, using NGC
      7410. The method is very similar to the \ion{Ca}{2} fit shown in
      Figure \ref{fig:CaTFit}.  {\bf Top:} A spline function models
      the continuum, using the \citet{men15} SSP galaxy templates to
      locate ``clean'' continuum points close to the lines of
      interest.  Real data is shown in blue, while the SSP template
      (re-scaled to match the flux level of the real data) is shown in
      green.  The knot-points of the spline model are represented by
      black squares and the actual spline fits to the data and
      template appear as a black lines. {\bf Middle:} The continuum is
      removed and each line is modeled with a Gaussian to measure an
      equivalent width.  The redshift and $\sigma$-value of each
      Gaussian is fixed (based on the results of the \ion{Ca}{2} fit),
      but the amplitudes are free to vary.  The best-fit model,
      measuring all lines simultaneously, is shown in red.  {\bf
        Bottom:} Same as the middle panel, but individual line
      components are shown.  The colors of each line component are the
      same as those used in Figure
      \ref{fig:allLines}.  \label{fig:Dfit}}
    \end{center}
\end{figure}

After creating this new continuum model, we again normalize and invert
the spectrum, allowing us to fit and measure the dwarf-line EWs.  This
process is also quite similar to the \ion{Ca}{2} model, though we
eliminate the redshift parameter and fix the sigma value to match the
best-fit \ion{Ca}{2} value.  We optimize the new model (consisting of
a set of 18 Gaussian amplitudes) using the same $\chi^2$-minimization
algorithm, and then re-run the process with Poisson errors to obtain
uncertainties.  Finally, we recreate all of the best-fit models and
sum up the model fluxes to obtain EW values for each line.  This
process is repeated for every galaxy in the sample, an example of
which is shown in Figure \ref{fig:Dfit}.  The complete fitting results
are shown in Table \ref{tbl:modParams}.

\begin{figure*}
  \includegraphics[width=17.6cm]{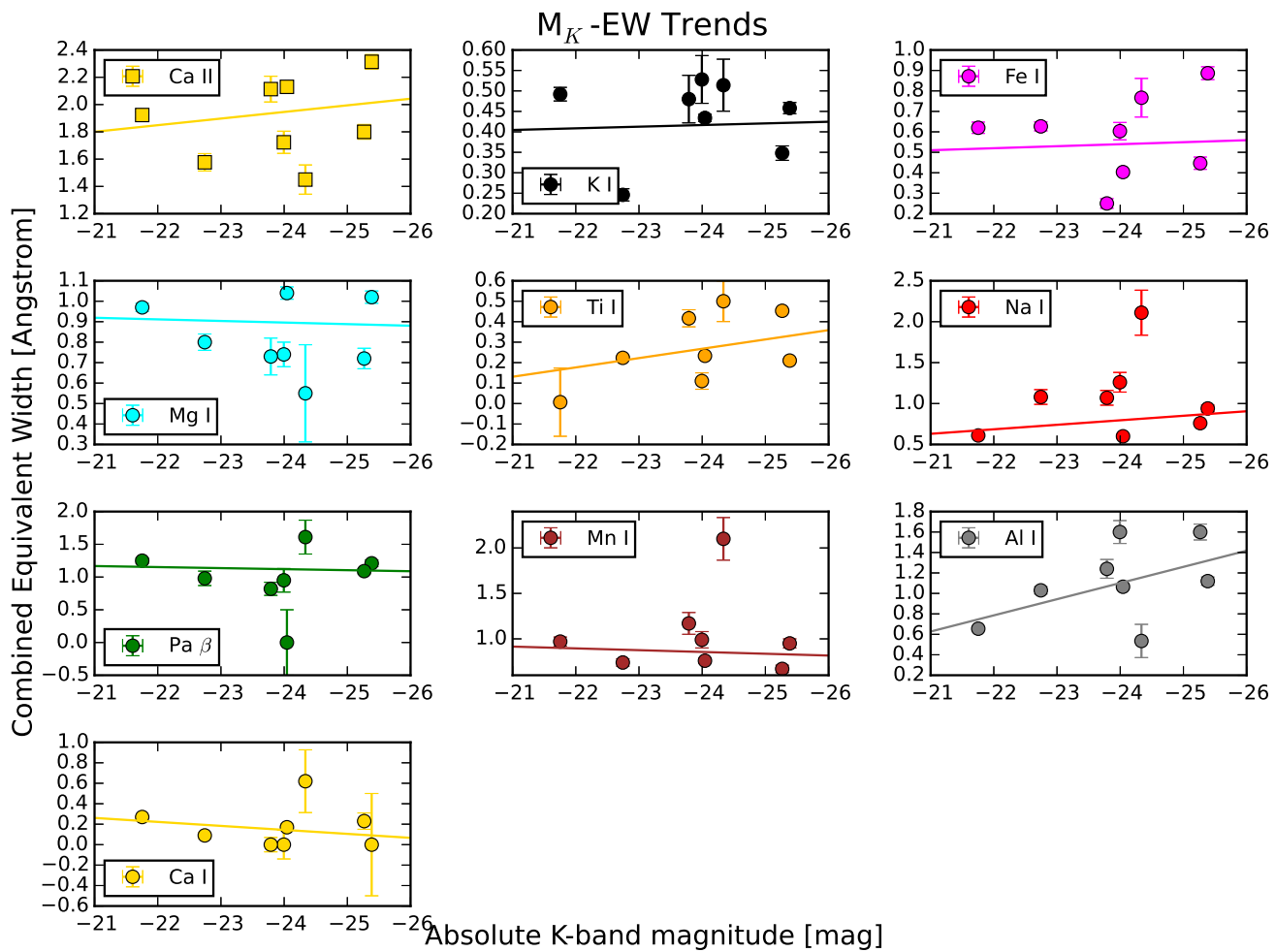}
  \caption{Measured equivalent width (EW) values for the metal lines
    used in our sample, as a function of K-band luminosity.  Elements
    which have multiple line features (e.g., \ion{Ca}{2} and
    \ion{K}{1}) are summed together to reduce noise.  A best-fit trend
    line for each element is also shown, given by the equation EW =
    $m_{\rm Mag} ~\times$ M$_{\rm K}$ + $b_{\rm Mag}$.  Fit parameters
    for each line can be found in Table \ref{tbl:TrendParams}.  While
    some elements do show a slight correlation between luminosity and
    EW, these trends tend to be weak and highly sensitive to
    scatter.\label{fig:magTrend}}
\end{figure*}

\begin{figure*}
  \includegraphics[width=17.6cm]{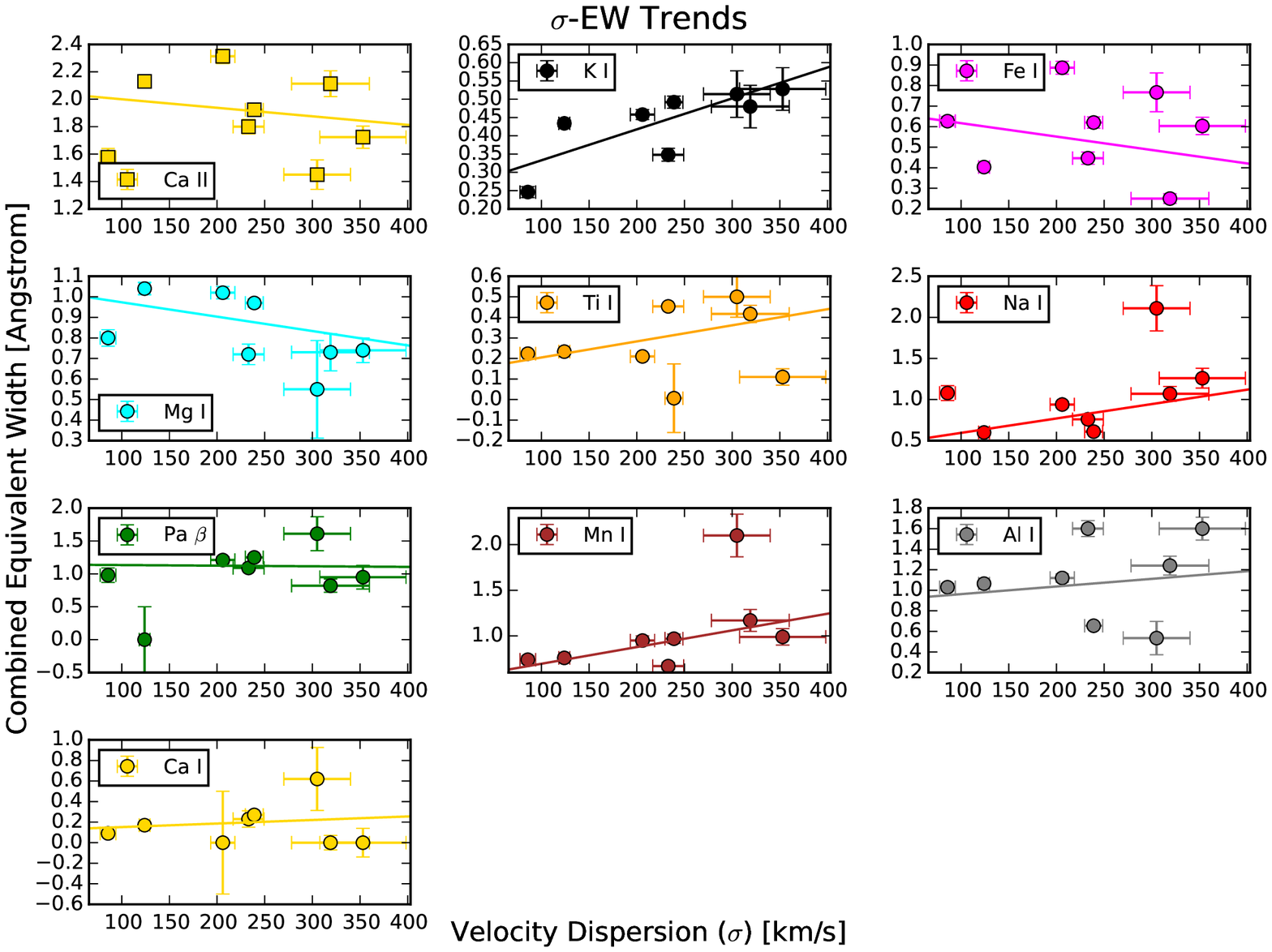}
  \caption{Same as Figure \ref{fig:magTrend}, but with velocity
    dispersion ($\sigma$).  Here, the trends are considerably more
    significant.  As expected, giant-sensitive features like
    \ion{Ca}{2} show a negative correlation between $\sigma$ and EW,
    while dwarf-sensitive features have a positive correlation.  Here,
    the equation of the trend line is given by EW = $m_{\sigma}\times
    \sigma$ + $b_{\sigma}$, and fit parameters are again presented in
    Table \ref{tbl:TrendParams}.\label{fig:sigTrend}}
\end{figure*}

\section{Analysis}

Trends between EW measurements and galaxy parameters (such as age,
mass, and environment) are caused by a variety of factors.  While many
are due to variations in chemical abundance, stellar age, and
metallicity, other effects can be caused by changes in the IMF.  To
find evidence of a variable IMF, we first look for trends between the
EW data and physical properties of the host galaxies, then compare the
results to theoretical models in order to account for the other, more
obvious factors (Section \ref{Models}).

Total mass, in particular, is an interesting quantity since previous
studies have demonstrated correlations between IMF slope and galaxy
mass (e.g., \citet{dut12,spi12,fer15,mar15a,spi15}).  Though there are
potentially many ways to measure mass, we focus here on two
complementary methods: In the first case, we measure the K-band
luminosity of each galaxy, assuming it to be a robust tracer of
stellar mass \citep[e.g.,][]{kau98} In the second case, we take the
best-fit velocity dispersion (determined from the pPXF fit) as the
mass proxy, since it directly probes the galaxy's central potential
well.

We then fit trend lines between mass and EW for each element (both
\ion{Ca}{2} and the Dwarf lines), using a standard least-squares
minimization technique that accounts for measurement errors in both
directions.  In instances where an element has more than one observed
absorption feature (i.e., \ion{Ca}{2}, \ion{K}{1}, \ion{Fe}{1},
\ion{Ti}{1}, and \ion{Al}{1}), we sum all lines together to create
a ``master'' EW, in order to reduce noise.  Figure \ref{fig:magTrend}
shows the results of the Luminosity-EW fit over the eight galaxies in
our sample.  While trends do exist for certain elements (e.g.,
\ion{Na}{1} and \ion{Al}{1}), the slopes are often weak and possibly
driven by one or two extreme outliers.  This fact, coupled with noisy
data and a small sample size makes it difficult to make a definitive
statement on the relationship between luminosity and EW.

On the other hand, the trends between EW and velocity dispersion
($\sigma$) (Figure \ref{fig:sigTrend}) are stronger, with some
elements (\ion{K}{1},\ion{Na}{1}, and \ion{Mn}{1}) having a clear
positive (EW-$\sigma$) correlation and others
(\ion{Ca}{2},\ion{Mg}{1}, and Pa-$\beta$) a negative one.  Following
\citet{van12}, a positive $\sigma$-EW correlation suggests that a line
is dwarf-sensitive, while a negative correlation suggests it is
giant-sensitive.  Thus, a negative slope for \ion{Ca}{2} (a known
giant-sensitive line) is expected, as are positive slopes for
\ion{K}{1},\ion{Na}{1}, and \ion{Mn}{1}.  The negative slopes for the
other lines could imply that they are giant-sensitive as well.
However, we stress that a positive (or negative) EW-$\sigma$ trend
alone cannot fully address IMF variability, and comparisons to
theoretical models accounting for age and metallicity are better
suited to answer this question.  Regardless, given that the
EW-$\sigma$ correlation is nominally stronger than that of
EW-Luminosity (in agreement with previous work by \citealt{gra09}), we
choose to use $\sigma$ as our mass tracer (and hence IMF tracer) for
the remainder of this work.  To do this, though, we will need to
compare our results to known IMF-sensitive models (e.g.,
\citealt{con10a}).

\section{Models}
\label{Models}

To probe IMF-dependency using models, we have written a high
resolution post-processor for the Flexible Stellar Population
Synthesis (FSPS) code base \citep{con09,con10b}.  We begin by running
FSPS normally, specifying metallicity ($Z$) and IMF parameters in the
usual way without changing the built-in libraries of evolutionary
tracks.  There are only $\sim$ 30 wavelength points in in the range
between 11500 $\AA$ and 13500 $\AA$ in FSPS, however, and we need
hundreds to probe at the FIRE resolution.  Therefore, our
post-processor substitutes the M star atmosphere fluxes of
\citet{all12} for the built in flux libraries of FSPS at effective
temperatures below 7000 K. This approach has the advantage that the
synthesis uses precise values of temperature, surface gravity, and
metallicity sampled as finely as necessary.  At higher temperatures we
simply interpolate the low resolution fluxes.

As previously mentioned, the principal parameters controlling a
galaxy's stellar spectrum are expected to be metallicity, star
formation history, IMF, and detailed chemical composition. A useful
parametrization of elemental composition is the $\alpha$/Fe ratio, as
supernova yields lead to such a scaling. A full investigation of early
type galaxy stellar spectra would employ techniques such as principal
component analysis to prioritize a moderate set of parameters, such as
metallicity, age, IMF slope and $\alpha$/Fe ratio. Such an
investigation is beyond the scope of the present work with its small
sample, and we instead focus only on first-order variations in IMF
slope. However, our library of synthetic spectra will permit
$\alpha$/Fe ratio and age to be investigated, when a large S/N
$\approx$ 100 sample becomes available. The present work is a
pathfinder for the telescope resources, both optical and infrared,
required for that study.

\begin{figure}
  \includegraphics[width=8.8cm]{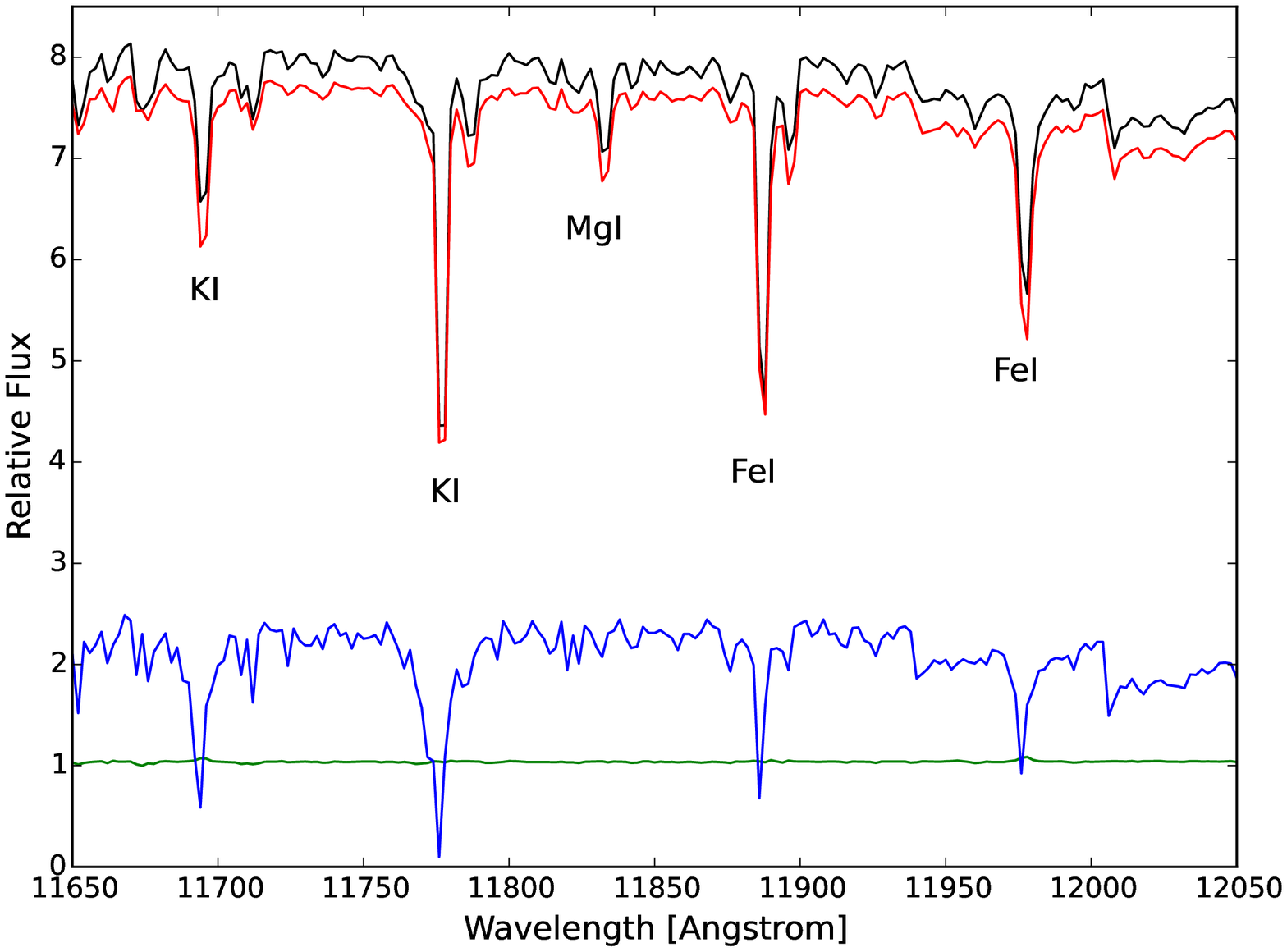}
  \includegraphics[width=8.8cm]{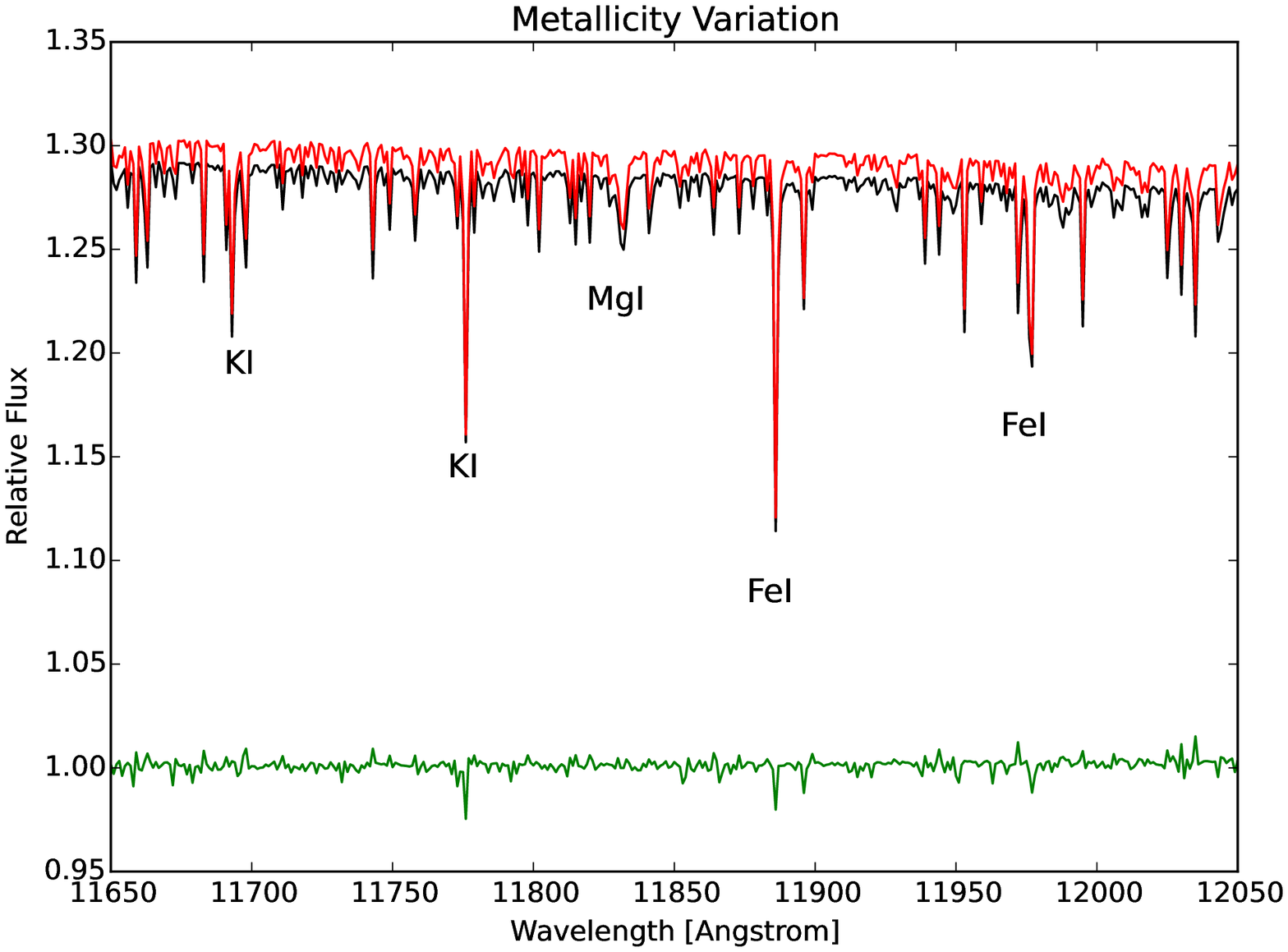}
  \includegraphics[width=8.8cm]{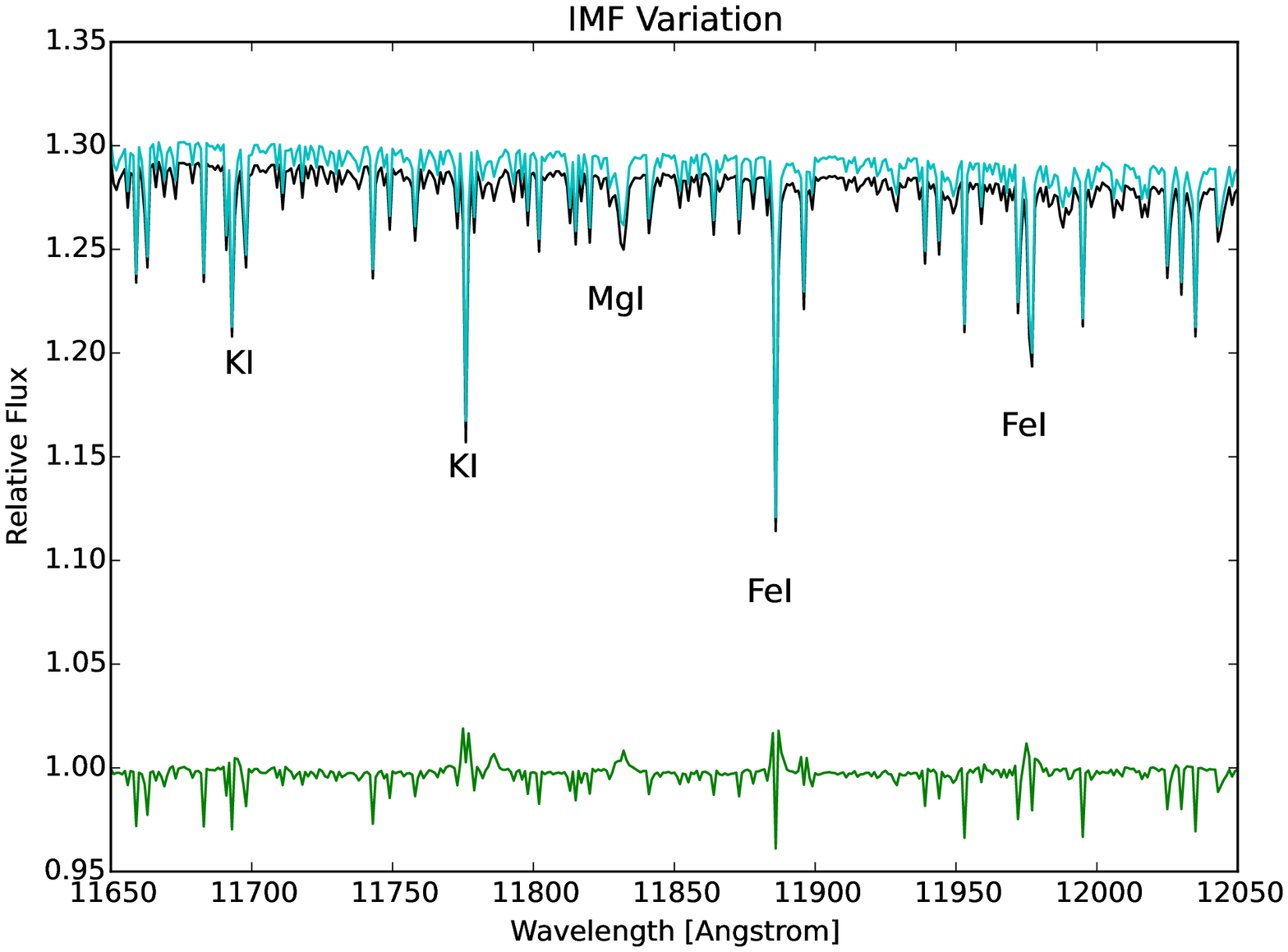}
\caption{{\bf Top:} Combined \citet{all12} model atmosphere spectra
  for 3000 K + 3100 K, $\log$(g) = 5. Broadening of 50 km s$^{-1}$ has
  been added.  The black spectrum is the model for solar metallicity
  and the red one for [M/H] = --0.5 dex.  These are ratioed in the
  green spectrum. The blue spectrum is for effective temperature 2500
  K.  {\bf Middle:} SSP models generated from the \citet{all12}
  spectra, assuming a Salpeter IMF and a stellar age of 10 Gyr.  Line
  colors are the same as in the top panel.  We see that a changing
  metallicity has an effect on EW values, though this effect is small.
  {\bf Bottom:} Similar SSP model comparison, but the low-metallicity
  model has been replaced by a solar-metallicity model with a Chabrier
  IMF (cyan line).  Comparing the ratio lines of the middle and bottom
  panels, we see a larger variance in the IMF test compared to the
  metallicity test, suggesting that a varying IMF may play a larger
  role in the observed EW trends.  We note, however that both effects
  are still small ($<$ 5\% for nearly all lines of interest).  In all
  panels, the positions of some metal lines used in this work are
  highlighted for reference. \label{fig:ModSpec}}
\end{figure}

Spectra for stellar models at Dwarf-star temperatures ($\sim$ 3100 K)
are shown in the top panel of Figure \ref{fig:ModSpec}.  Along with a
model with solar metallicity (black line), an additional variant with
$Z = Z_\odot/3$ ([M/H] = --0.5 dex) is shown in red.  A
low-temperature (2500 K) model, also at solar metallicity, is
presented for reference (blue line).  Looking at the ratio between the
black and red lines (green line), we see that for M dwarf temperatures
the metallicity dependence of the main sequence flux is weak. This is
due to the line opacity and continuum opacity scaling in a similar way
with metallicity. The \ion{K}{1} line, however, becomes very strong at
L dwarf temperatures.  Folding this information into an integrated
galaxy spectrum, we show the final SSP models for both the $Z =
Z_{\odot}$ and $Z = Z_\odot/3$ cases, assuming a Salpeter IMF, in the
middle panel of Figure \ref{fig:ModSpec}.  Here we see a slightly more
significant dependence on metallicity when main sequence flux is
included.  However, the effect is still relatively small and does not
account for all of the EW variation seen in the data, leaving room for
potential IMF effects.  We probe for these effects directly in the
bottom panel of Figure \ref{fig:ModSpec}, where we replace the
low-metallicity Salpeter model with a solar-metallicity Chabrier model
(cyan line).  While the overall ratio is still close to unity, there
is a noticeably larger variance between the depths of the metal lines
in this case, suggesting that IMF may play a larger role in
determining their EWs.  However, while the IMF variation is larger
than the metallicity variation, both effects are still weak: in all
but the most extreme cases, the typical change in line strength from
model to model is less than 5\%.

\begin{figure}
  \includegraphics[width=8.8cm]{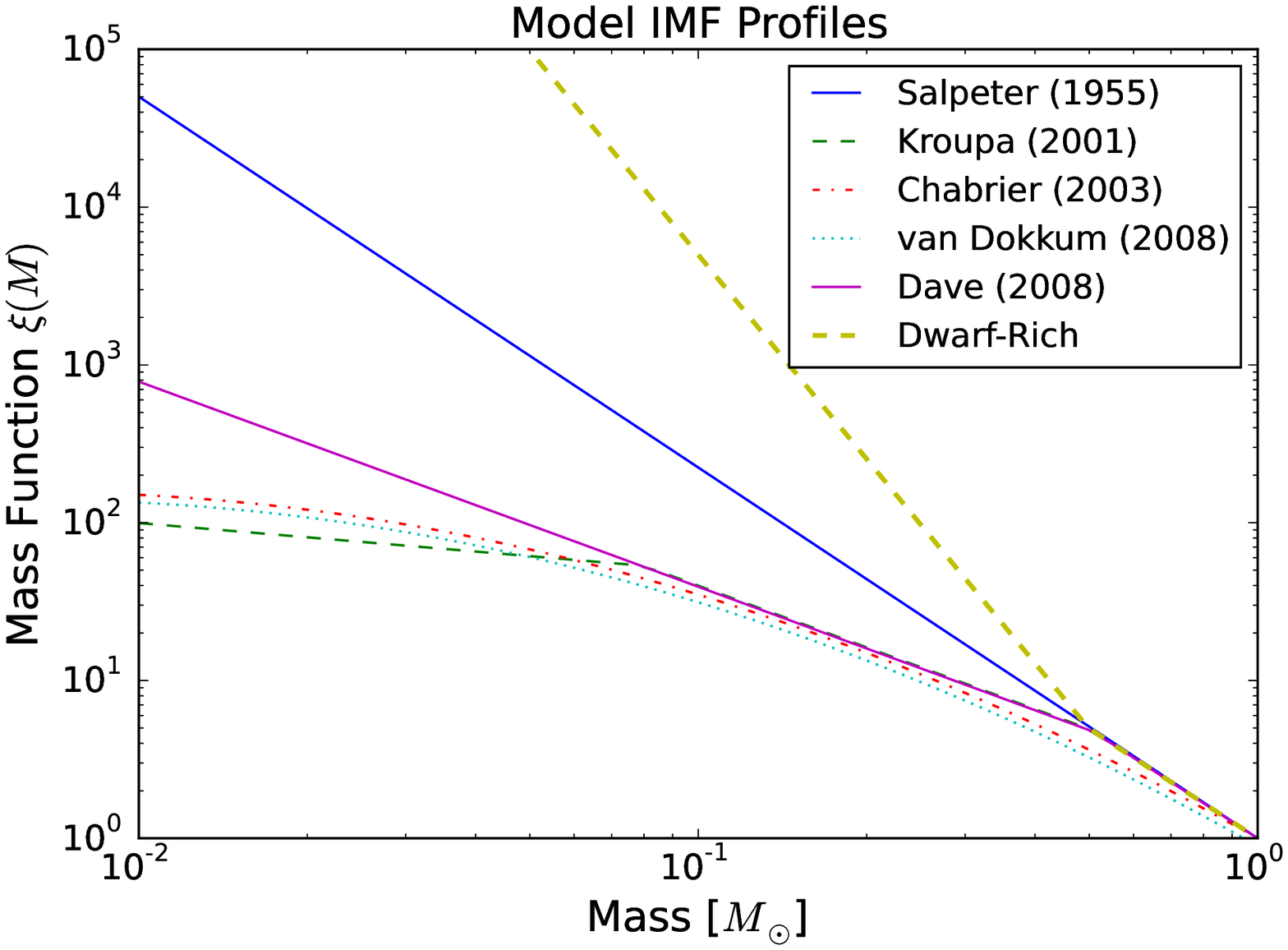}
  \caption{Mass profiles of each of the IMF models used in this work.
    All profiles are normalized at 1 $M_{\odot}$.  Using each IMF as
    an input parameter in SSP modeling, we are able to measure the
    effects of a variable IMF on the EWs of our target metal lines.
    The results of this modeling are presented in Table
    \ref{tbl:ModLines}. \label{fig:IMFs}}
\end{figure}

Using a series of these synthetic spectra, each generated with a
different model IMF, we measure theoretical EW values for both the
\ion {Ca}{2} triplet and the M-dwarf metal lines, which we compare to
the observed data.  To be as consistent as possible, we measure model
EWs using the technique outlined in Section \ref{Measurements}, and
like the observed data, we smooth the models to a common velocity
dispersion $\sigma = 350$ km s$^{-1}$.  EWs for all lines are recorded
in Table \ref{tbl:ModLines}.  For our choice of IMF, we use all five
IMF codes of \citet{con10b}, namely: \citet{sal55}, \citet{kro01},
\citet{cha03}, \citet{van08}, and \citet{dav08}, along with an
additional dwarf-rich variant.  The dwarf-rich model takes the form of
a broken power-law, with a steep slope ($\alpha$ = 4.3) at the
low-mass end ($0.08 M_\odot < M_{\rm star} < 0.5 M_\odot$), and a
shallower slope ($\alpha$ = 2.3) at higher masses ($M_{\rm star} > 0.5
M_\odot$.)  (For comparison, \citet{sal55} has a single power-law
slope $\alpha$ = 2.35). While this low-mass slope parametrization is
considerably steeper than the $\alpha \sim 3$ slope observed in other
early-type galaxies \citep[e.g.,][]{spi14,con17}, this is an
intentional choice: by including a model with a super-abundance of
dwarf stars, we have a larger baseline for comparing models, allowing
us to more easily understand and isolate the IMF's role in line
strength variation. A graphical representation of each IMF model can
be seen in Figure \ref{fig:IMFs}.

In the table, we measure EWs for both solar and sub-solar metallicity
parameters, though we limit stellar age to single-burst populations
between 10 and 13 Gyr in order to keep the total parameter space
small.  These ages are expected to match well with the stellar ages of
our galaxy sample. The metallicity dependence of the EWs is visible in
the table, as are somewhat larger dependencies on stellar age and IMF.

\begin{figure*}
  \begin{center}
    \centerline{
      \includegraphics[width=6.3cm]{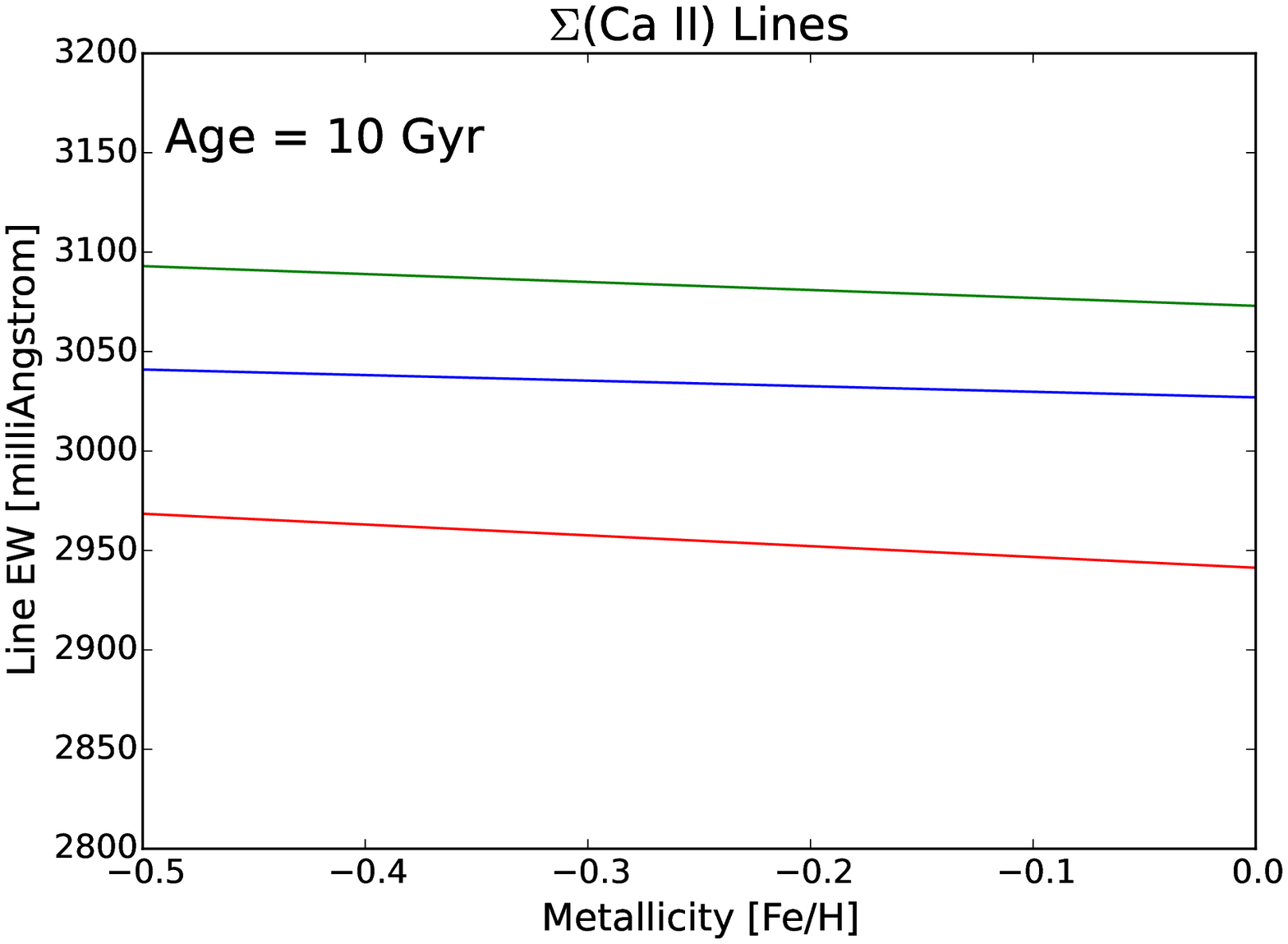}
      \includegraphics[width=6.3cm]{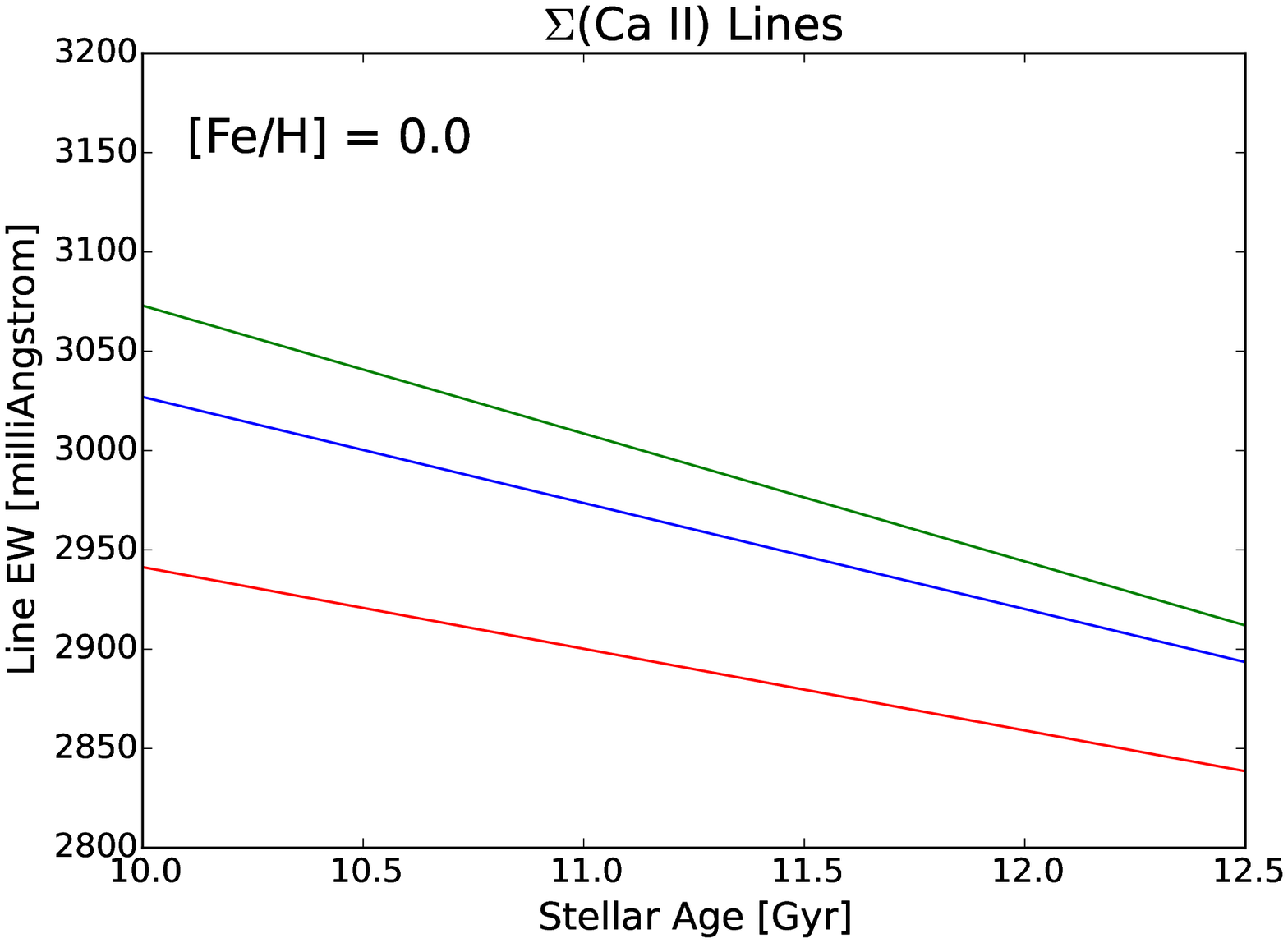}
    }
    \centerline{
      \includegraphics[width=6.3cm]{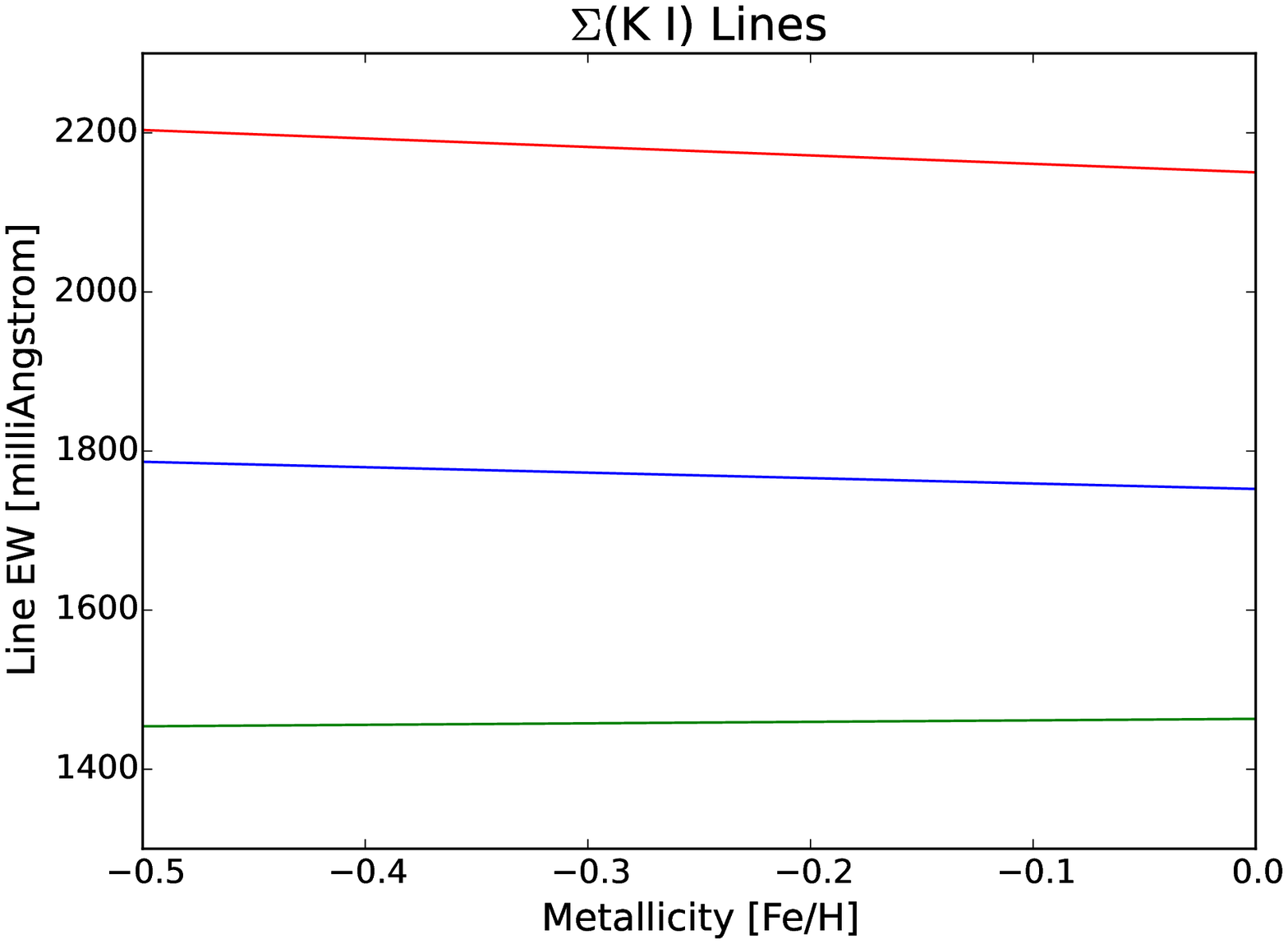}
      \includegraphics[width=6.3cm]{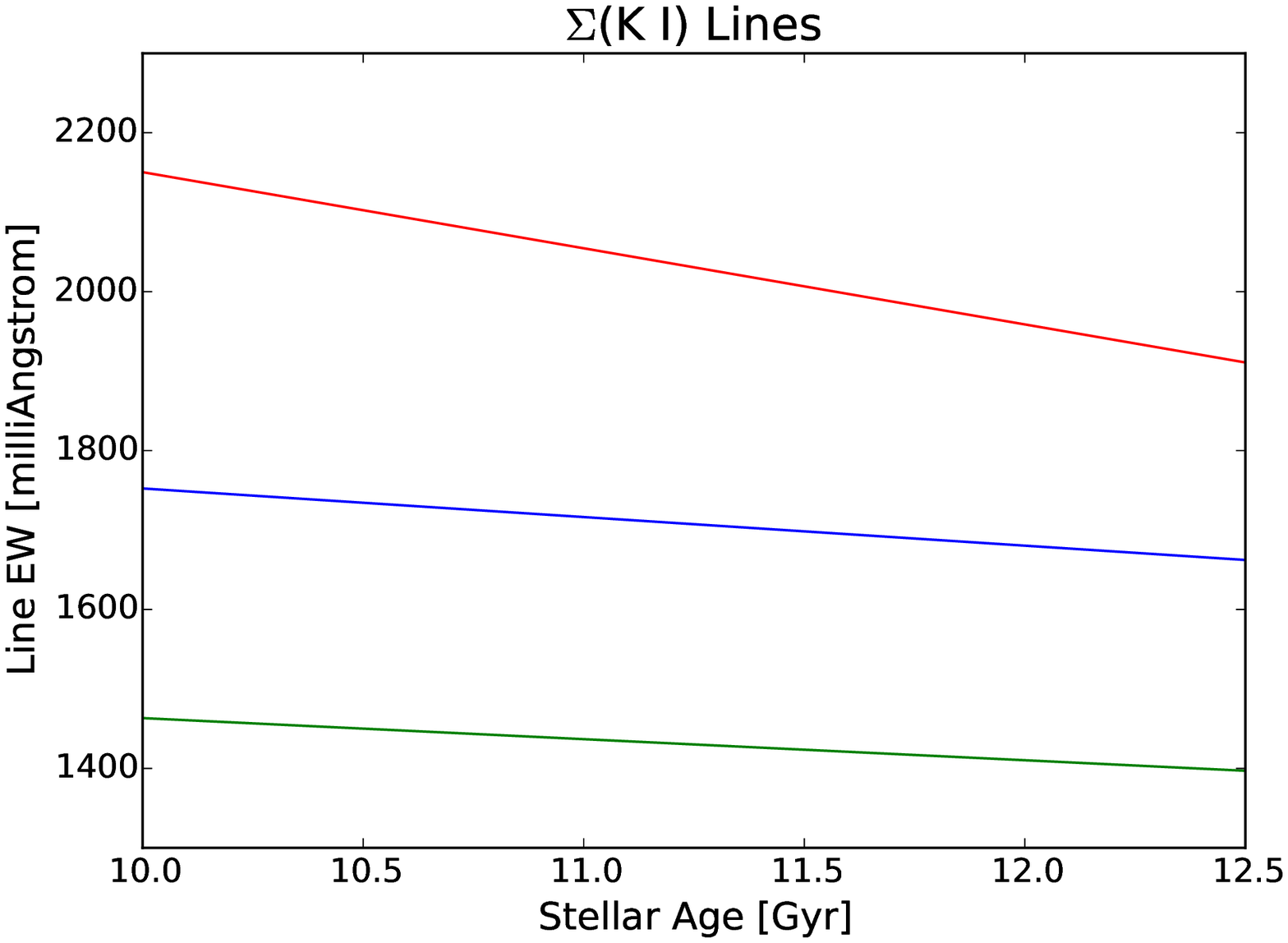}
    }
    \centerline{
      \includegraphics[width=6.3cm]{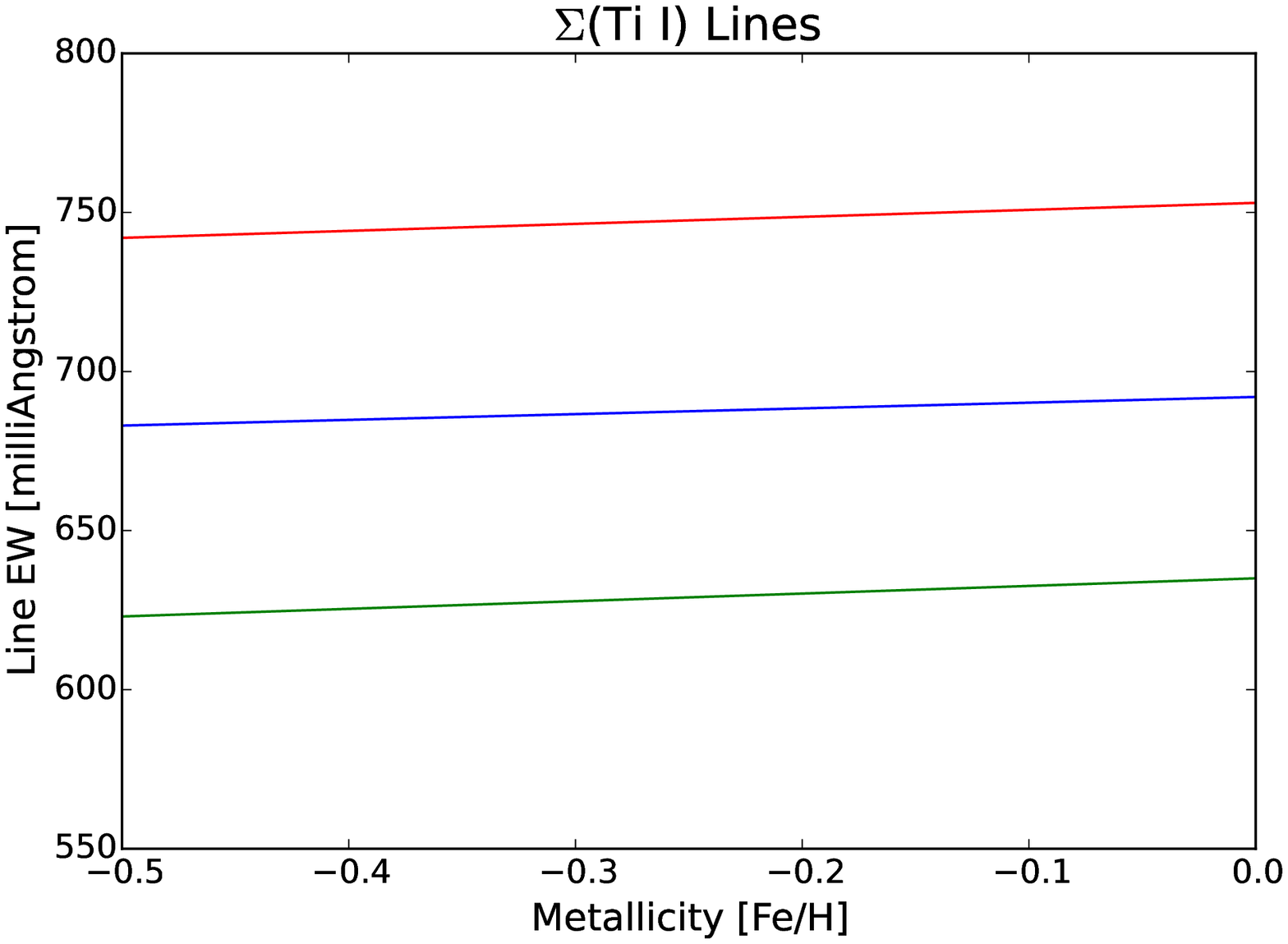}
      \includegraphics[width=6.3cm]{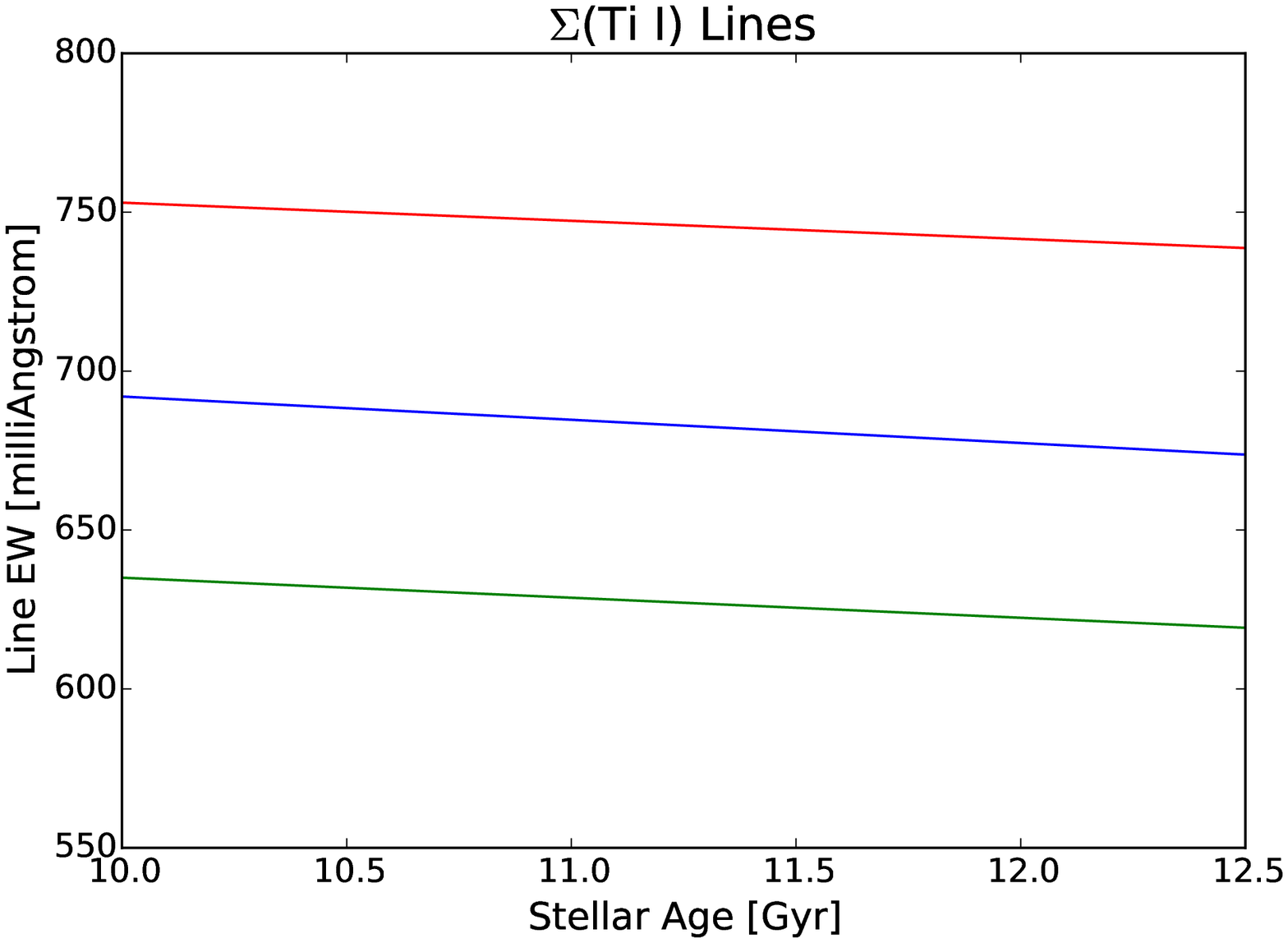}
    }
    \centerline{
      \includegraphics[width=6.3cm]{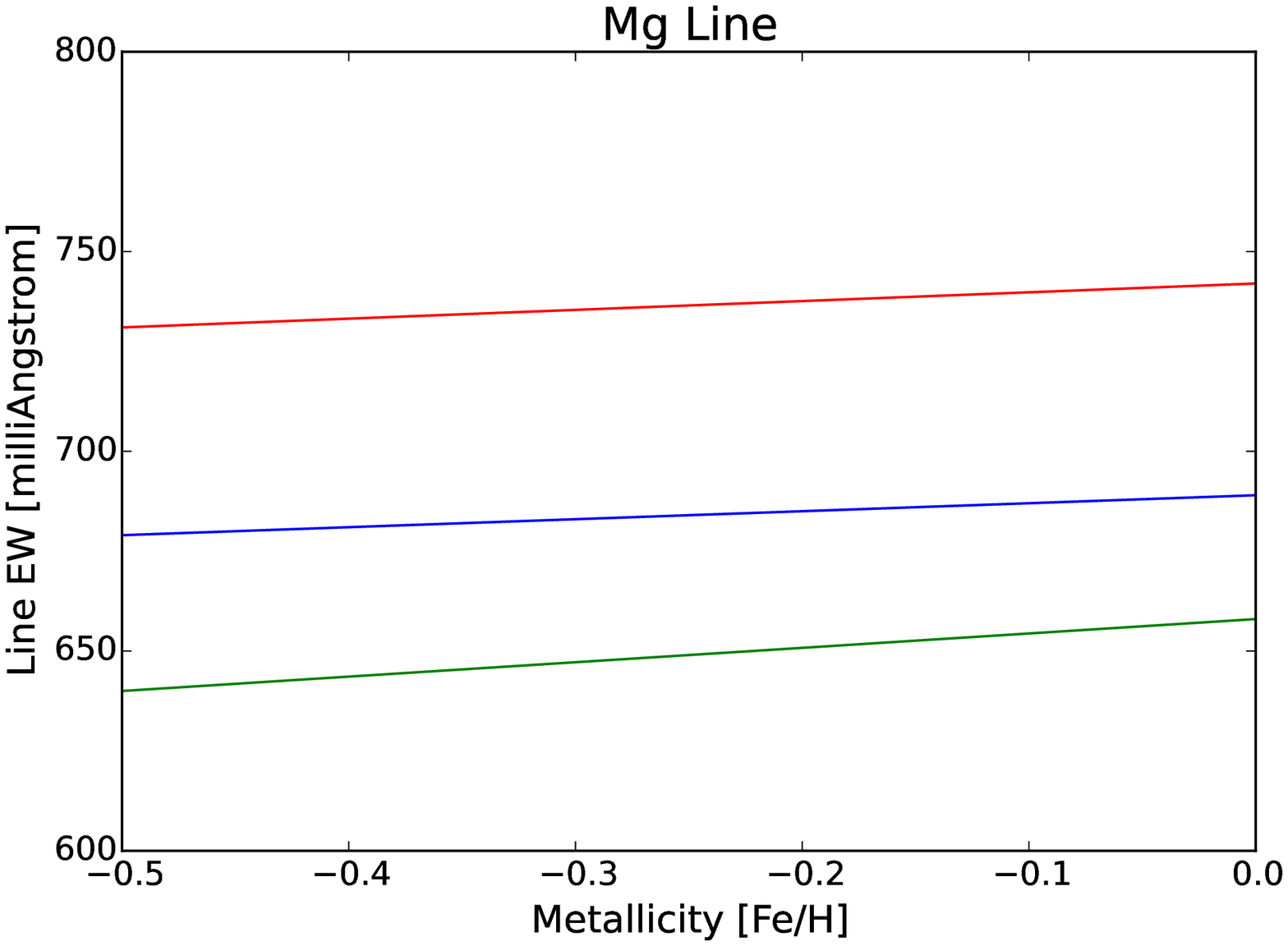}
      \includegraphics[width=6.3cm]{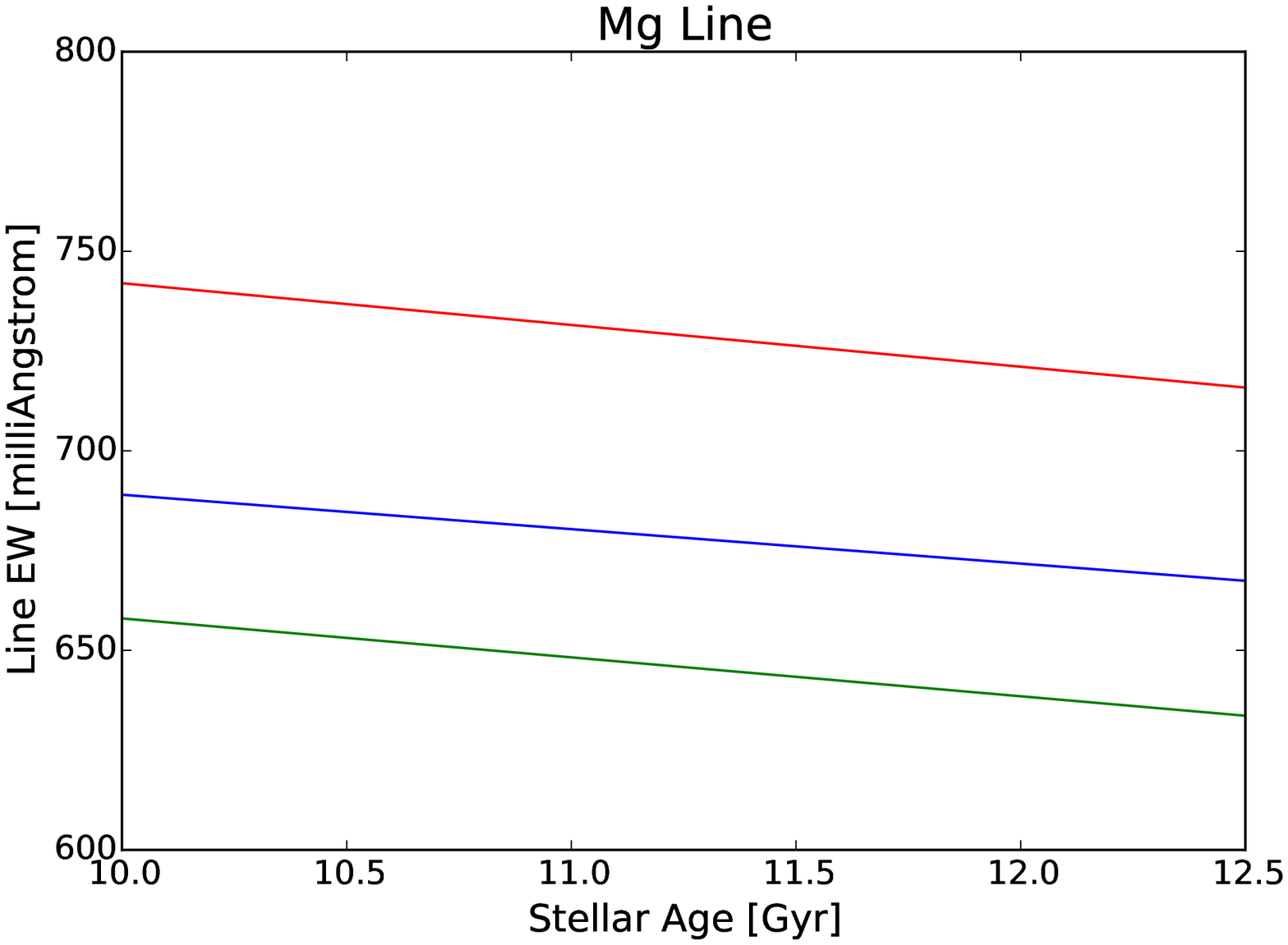}
    }
    \centerline{
      \includegraphics[width=6.3cm]{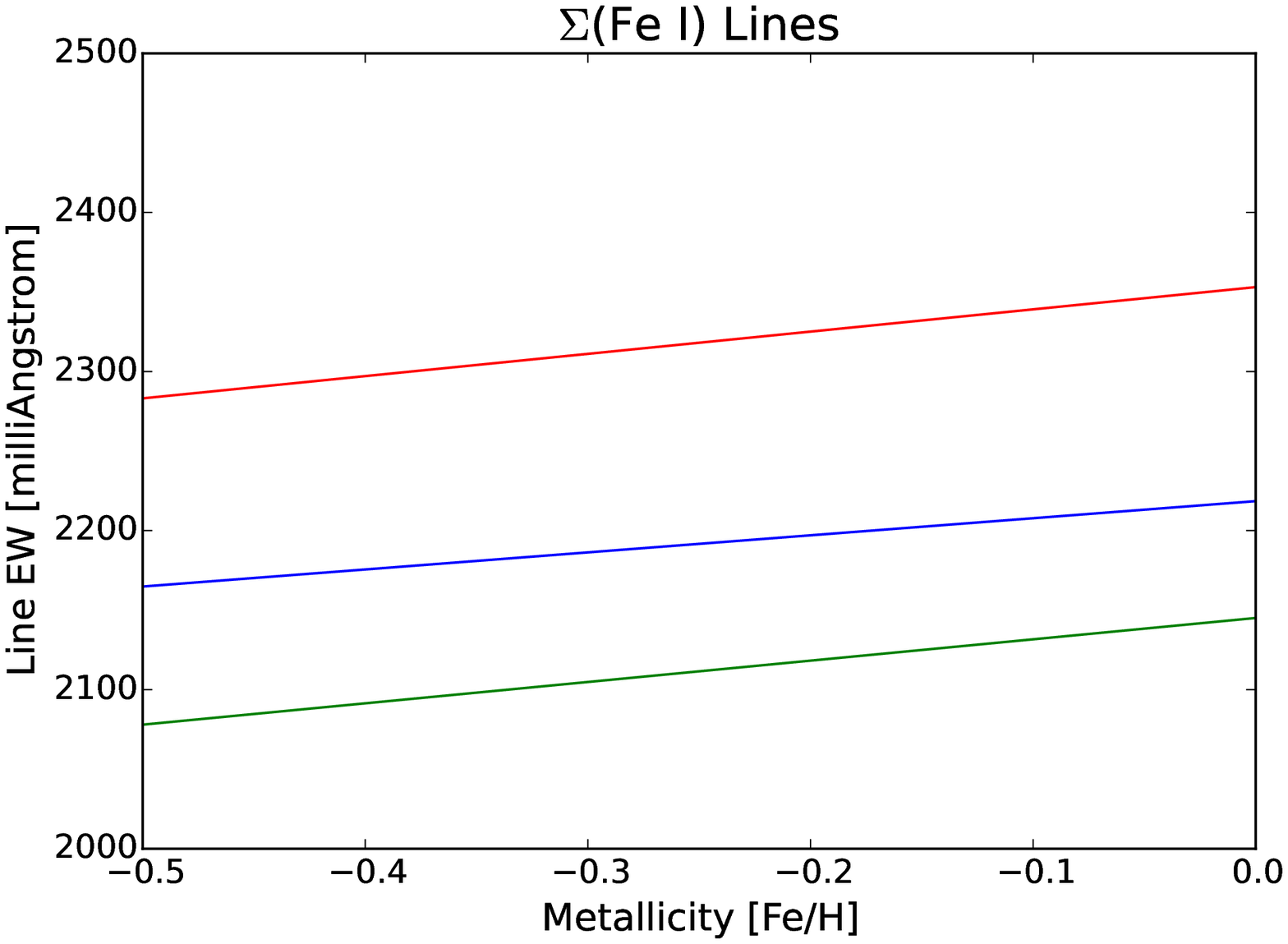}
      \includegraphics[width=6.3cm]{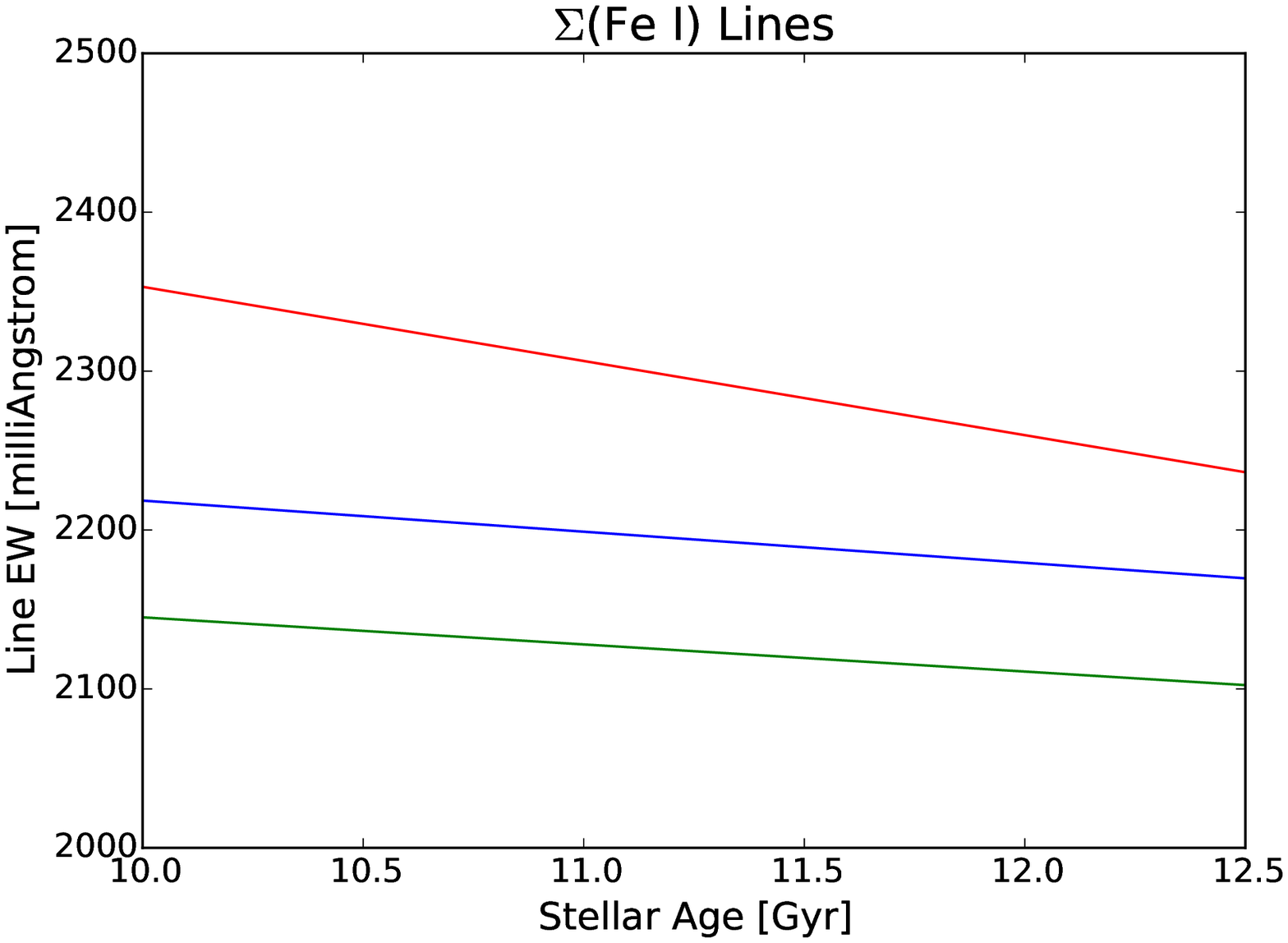}
    }
    \caption{Variations in line EW as a function of metallicity and
      stellar age.  Overall, we see that metallicity effects (left
      column) tend to be weaker than age effects (right column), but
      both play as important role.  To avoid parameter degeneracies,
      the metallicity models are held at a fixed stellar age of 10
      Gyr, while the age models are held at a fixed metallicity [Fe/H]
      = 0.  To account for the IMF, we plot three different IMF codes:
      Salpeter (blue), Chabrier (green) and the Dwarf-Rich variant
      (red).  We do see significant offsets between IMF models, which
      are larger than the variations due to metallicity and at least
      comparable to the variations due to stellar age.  While the
      magnitude of the variations change with different IMF models,
      the general trend (positive or negative) of each element remains
      the same. \label{fig:TheoryTrends}}
  \end{center}
\end{figure*}

We highlight the effects of metallicity and stellar age on various
line strengths in Figure \ref{fig:TheoryTrends}.  From the plot, we
see slight trends between line EW and metallicity, though the effect
is minimal and EW remains largely flat between sub-solar and solar
models.  As a sanity check, we do see a mild increase in \ion{Fe}{1}
EW with increasing metallicity, which is to be expected given the
iron-based metallicity variations we employ.  A stronger effect can be
seen with stellar age, where the older models show consistently
smaller EWs.  To account for IMF effects we include three different
IMF codes: Chabrier (green lines), Salpeter (blue lines), and the
Dwarf-Rich variant (red lines), which span our full range of low-mass
IMF slopes.  Overall, we see that the IMF is also an important factor,
having a much stronger effect than metallicity and at least as
comparable an effect as stellar age.

While there is some tension between our theoretical trends and those
seen in other works (in particular, \citealt{cen03} see an essentially
flat EW trend with stellar age as opposed to our decreasing trend),
this may not be entirely unexpected.  Specifically, in evolutionary
population synthesis there are significant uncertainties in the
Hertzsprung-Russell diagram location of the core helium stage in very
old populations and the double shell burning stage (asymptotic giant
branch; AGB) in intermediate age populations. Although neither of
these is the dominant light source at one micron, they are responsible
for of order 10\% of the light. We have simply used the default
prescriptions in FSPS in the present paper but emphasize the caveat
that line strength and age predictions in our current models are
uncertain at levels of at least a few percent. These will need to be
investigated further in our ongoing work with FSPS.  Our model
predictions regarding IMF dependence are on a firmer footing, as they
are primarily influenced by the red giant branch and lower main
sequence ratio. By accounting for age and metallicity effects,
however, we are able to better measure the intrinsic variations due to
the IMF.

To demonstrate an IMF dependence that is less sensitive to metallicity
effects, we plot the ratio of \ion{K}{1} to \ion{Mg}{1} as a function
of $\sigma$.  We choose \ion{K}{1} for the numerator because it is a
strong dwarf/giant discriminator, and \ion{Mg}{1} for the denominator
because of its strong dependence on mass.  Potassium is mostly ionized
in the atmospheres of M dwarfs, but magnesium is not.  As in the
previous section, we sum the individual \ion{K}{1} lines together to
reduce noise, resulting in the ratio:

\begin{equation}
  R = \frac{\sum \rm EW(K~I)}{\rm EW(Mg~I)}
\end{equation}

\begin{figure*}
   \begin{center}
    \centerline{
      \includegraphics[width=15.6cm]{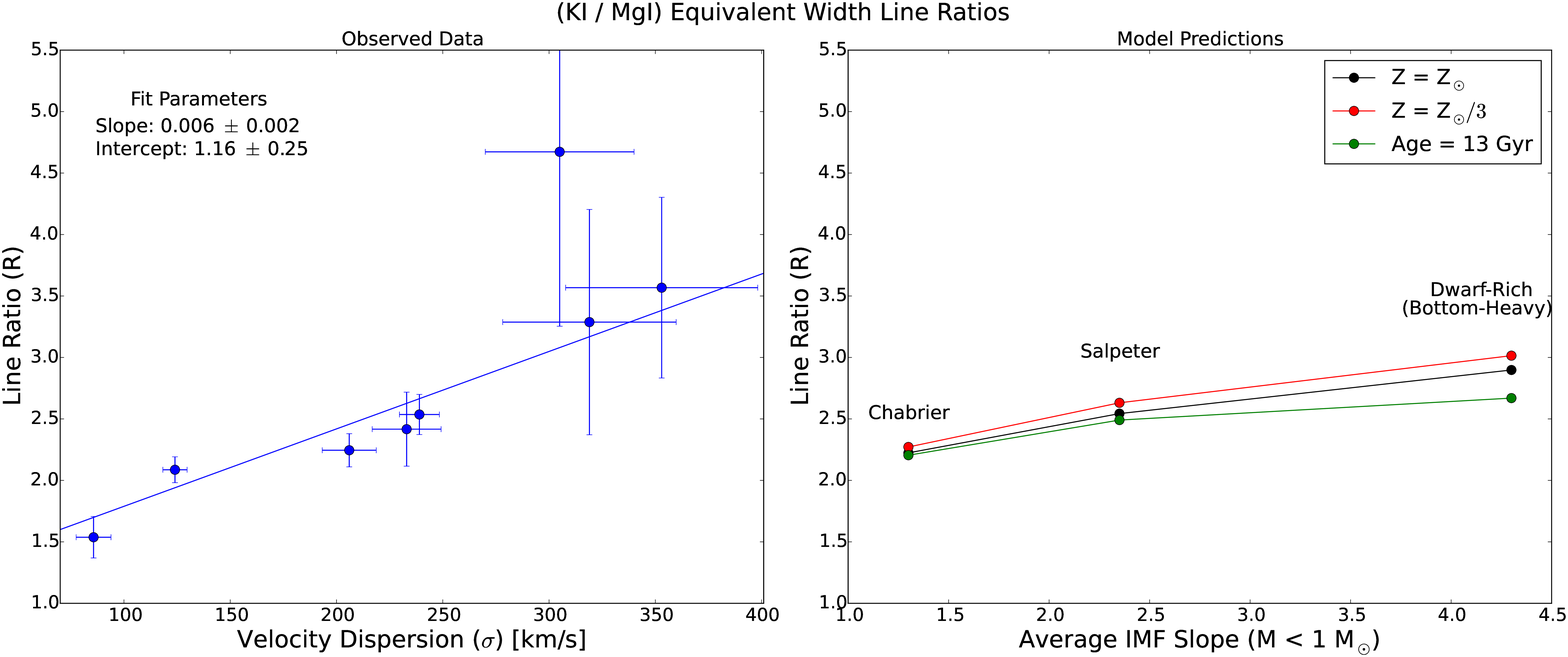}
    }
    \centerline{
      \includegraphics[width=15.6cm]{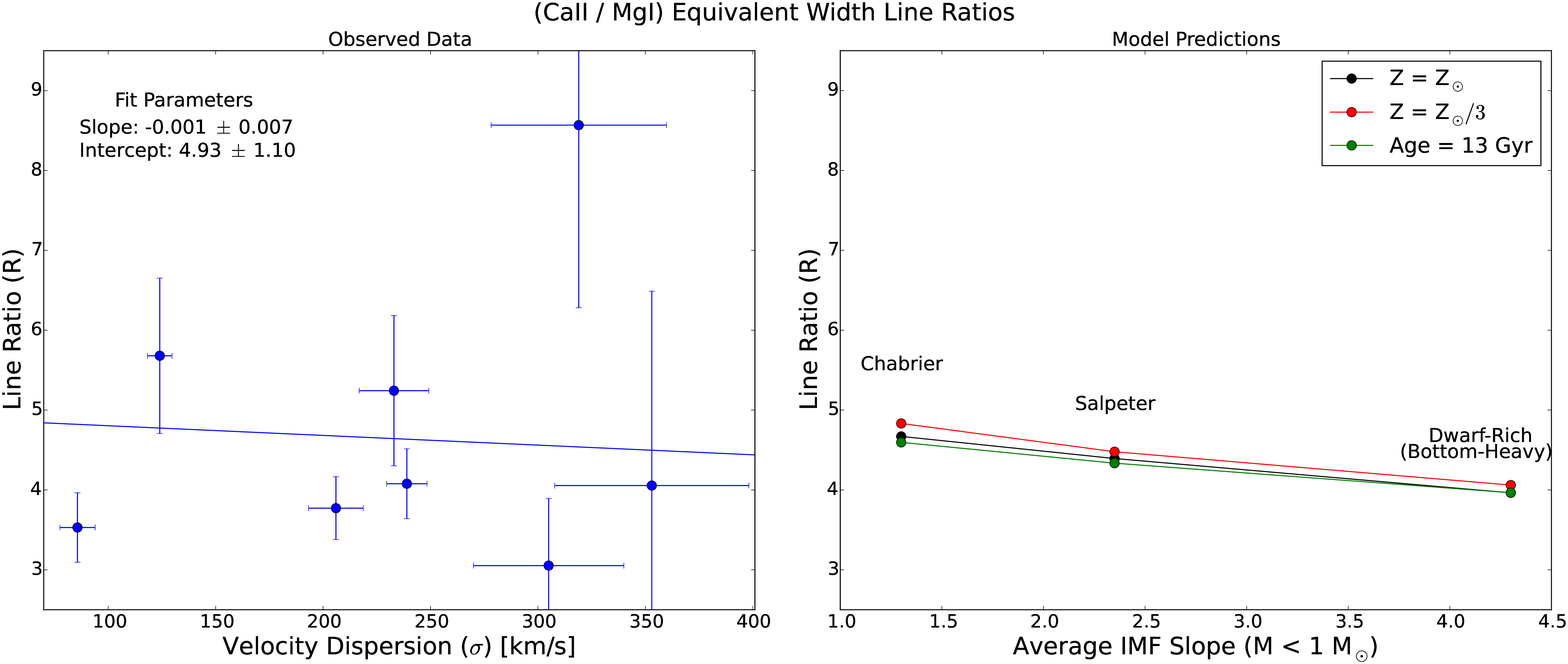}
    }
    \centerline{
      \includegraphics[width=15.6cm]{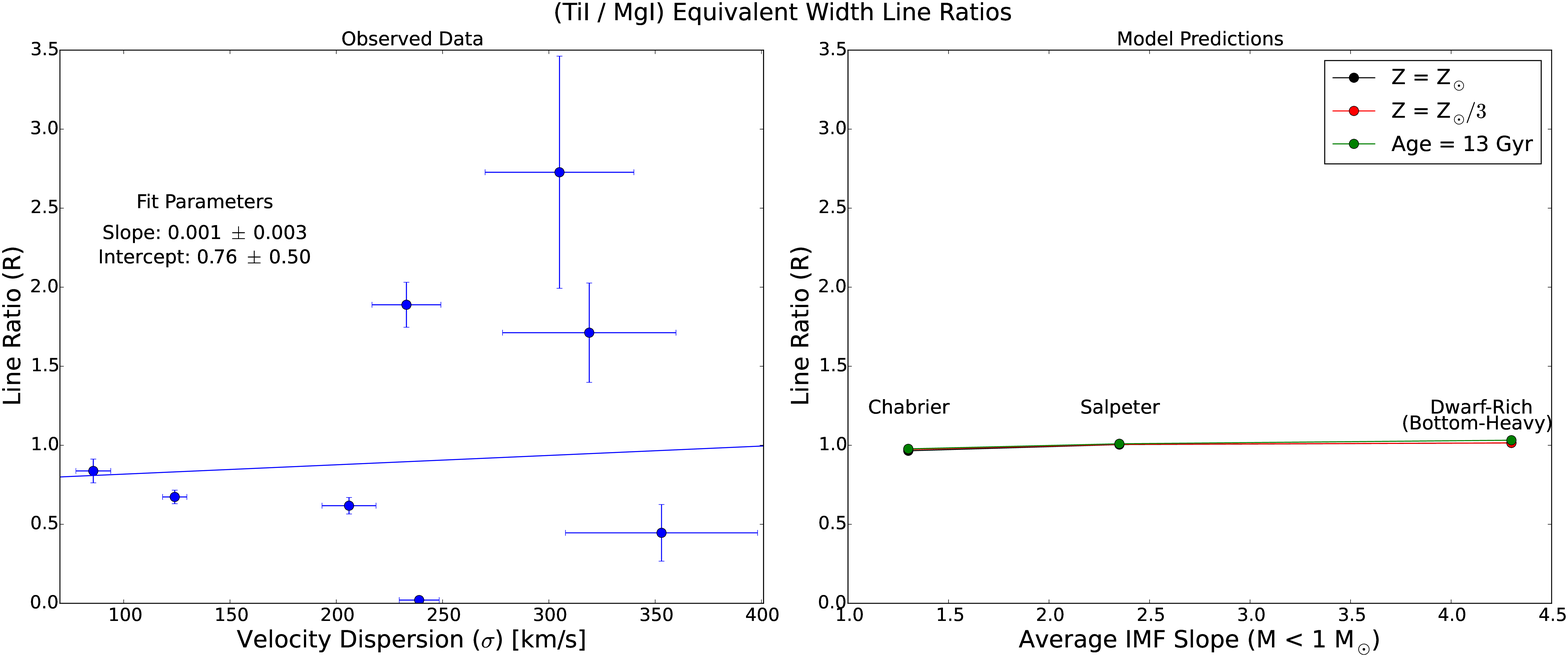}
    }
 \caption{{\bf Top Left:} Ratio of \ion{K}{1} to \ion{Mg}{1} line
   strengths as a function of velocity dispersion ($\sigma$).  Ratios
   for each galaxy in the sample are shown as blue circles.  A
   best-fit trend line can also be seen, highlighting a strong
   positive correlation between the two quantities.  {\bf Top Right:}
   Expected \ion{K}{1} / \ion{Mg}{1} line ratios from the models based
   on the \citet{con10b} FSPS package, using three different IMF
   slopes. The black line represents models with $Z = Z_\odot$
   metallicity, while the red line instead assumes a more metal-poor
   $Z = Z_\odot/3$ population.  The green line again
   assumes a solar metallicity, but is for a stellar age of 13 Gyr
   instead of 10 Gyr.  {\bf Lower Panels:} Similar line ratios using
   \ion{Ca}{2} and \ion{Ti}{1}, elements that are already known to be
   IMF sensitive.  In both of these cases, we do see a moderate trend
   in the data and an agreement with the models, however the slopes
   are not as significant and the scatter is
   larger. \label{fig:ModCompare}}
 \end{center}
\end{figure*}

The results, using our observed galaxy data, are shown in the upper
left panel of Figure \ref{fig:ModCompare}.  The best-fit trend line to
these data shows a significant correlation between line ratio and
$\sigma$, given by $R$ = (0.6 $\pm$ 0.2)$\sigma_{100}$ + (1.16 $\pm$
0.25), where $\sigma_{100}$ is $\sigma$ expressed in units of 100 km
s$^{-1}$.  We do note that scatter in the data points could alter the
slope, and the small sample size is subject to selection effects,
however, the overall form of the trend is stable.

In the upper right-hand panel of Figure \ref{fig:ModCompare} we plot
the expected \ion{K}{1} / \ion{Mg}{1} ratio for the theoretical
models, assuming various IMF slopes.  The black line assumes a stellar
population with solar metallicity, while the red line uses a more
metal poor ($Z_{\rm gal} = Z_{\odot}/3$) model.  The green line again
assumes solar metallicity, but uses an older stellar population (13
Gyr) than either the black or red models (10 Gyr.)  In all cases, we
see that steeper IMF slopes predict a larger line ratio.  While the
magnitude of this effect is dependent on metallicity and stellar age,
these changes are typically smaller than that of the IMF, diminishing
their importance in the final result.  While we do not explicitly test
the effects of chemical abundance, we note that in the case of
potassium \citet{tak09} demonstrated that stars show only a mild trend
in [K/Fe] abundance over the metallicity range populated by galaxies
with 100 $< \sigma <$ 350 km s$^{-1}$.  Because of this, potassium
abundance should not play a significant role in the EW trends we see.

By comparing the two panels of the figure, we see that the measured
data points have a similar shape to the model predictions, suggesting
that the IMF is indeed variable, with low-mass galaxies favoring a
Chabrier-like IMF and high-mass galaxies preferring a steeper,
dwarf-heavy population, in agreement with other studies.  Although the
model prediction shows a shallower trend than the observed slope, the
two plots are fully consistent with each other given the observational
errors.  

As a final comparison, we plot similar line ratios in the lower panels
of Figure \ref{fig:ModCompare}, however, here we substitute
\ion{Ca}{2} and \ion{Ti}{1}, two elements which are already known to
be sensitive to variations in the IMF.  Like the \ion{K}{1} ratio,
these elements also show a slight trend favoring a bottom-heavy IMF in
massive galaxies, though in both cases the correlation is less
significant and there is more scatter in the data.  Here again, the
theory and observation trends agree with each other within the
uncertainties.  Other element lines have even noisier data, leading to
less robust line fits.  This is likely due to our small sample size,
and future efforts with a larger galaxy sample will help to reduce
statistical uncertainty.  Regardless, these results, coupled with the
more significant \ion{K}{1} fit are promising, and do suggest some
form of IMF variability.

\section{Conclusions}
We have presented the first results of a study looking for possible
IMF-sensitive line features in the near-infrared ($\lambda$ = 1.1 --
1.3 $\mu$m.)  For this pilot program we use a sample of eight massive,
luminous ($L \sim 10L_*$), nearby galaxies, observing each with the
Magellan FIRE spectrograph in order to maximize SNR and spectral
coverage.  We identify candidate lines using dwarf-star line catalogs
\citep{mcl07,des12}, and also include the \ion{Ca}{2} triplet (a known
IMF-sensitive feature; \citet{van12}) as a calibrator.

After identifying the lines, we measure EWs from the FIRE spectra and
then look for correlations in the EW-mass plane.  To trace total mass
in each galaxy, we select two physical properties: K-band luminosity
(M$_{\rm K}$) and central velocity dispersion ($\sigma$).  While we do
detect mild correlations between EW and luminosity, the best-fit trend
lines are noisy and possibly driven by outliers.  Conversely, the
EW-$\sigma$ relationships are nominally tighter, with some elements
showing a strong positive (possibly Dwarf-sensitive) correlation
(\ion{K}{1},\ion{Na}{1},\ion{Mn}{1}), and others a negative (possibly
Giant-sensitive) correlation (\ion{Ca}{2}, \ion{Mg}{1}, Pa-$\beta$).

We probe the data for IMF-dependence by taking the EW ratio ($R$) of
\ion{K}{1} to \ion{Mg}{1} as a function of $\sigma$.  After fitting a
trend line to the data, we find a strong positive correlation given by
$R$ = (0.6 $\pm$ 0.2)$\sigma_{100}$ + (1.16 $\pm$ 0.25). Comparing
this trend to the expected $R$ value of theoretical models, this
suggests a changing IMF: from dwarf-poor (Chabrier) in low-mass
galaxies, to dwarf-rich in high-mass galaxies.  While we do note the
small sample size makes our data set susceptible to systematic bias
and selection effects, our results are promising, even in the face of
these limitations.  Future efforts, involving a larger sample of
galaxies, will likely improve our understanding of this phenomenon
even further.

\acknowledgements The authors wish to thank the referee for thoroughly
reading the manuscript and providing several useful comments which
greatly improved the quality of the paper.  Thanks to Rob Simcoe for
assistance with the FIRE pipeline. France Allard provided invaluable
advice on synthetic stellar spectra. We also thank Blesson Mathew and
Nidia Morrell for their contributions.  We appreciate the use of
Charlie Conroy's FSPS and Michele Cappellari's pPXF codes.  DJL
acknowledges support from the ERC starting grant 336736-CALENDS.
Parts of this research were conducted by the Australian Research
Council Centre of Excellence for All-sky Astrophysics (CAASTRO),
through project number CE110001020.  DAF thanks the ARC for financial
support via DP130100388.

\begin{turnpage}
\tabletypesize{\tiny}
\setlength{\tabcolsep}{0.02in}
\begin{deluxetable*}{ccccccccccccccccccccccccc}
\tablecaption{Best-fit model parameters and equivalent width values \label{tbl:modParams}}
\tablehead{
\colhead{Galaxy} & \colhead{$z$} & \colhead{$\sigma$} & \colhead{$\rm M_K$} &\colhead{\ion{Ca}{2}} & \colhead{\ion{Ca}{2}} &\colhead{\ion{Ca}{2}} & \colhead{\ion{K}{1}} & \colhead{\ion{K}{1}} & \colhead{\ion{K}{1}} & \colhead{\ion{Fe}{1}} & \colhead{\ion{Mg}{1}} & \colhead{\ion{Fe}{1}} & \colhead{\ion{Ti}{1}} & \colhead{\ion{Fe}{1}} & \colhead{\ion{K}{1}} & \colhead{\ion{K}{1}} & \colhead{\ion{Na}{1}} & \colhead{\ion{Ti}{1}} & \colhead{Pa-$\beta$} & \colhead{\ion{Ti}{1}} & \colhead{\ion{Mn}{1}} & \colhead{\ion{Al}{1}} & \colhead{\ion{Ca}{1}} & \colhead{\ion{Al}{1}}\\
\colhead{} & \colhead{} & \colhead{(km s$^{-1}$)} & \colhead{} & \colhead{$\lambda$8498} & \colhead{$\lambda$8542} & \colhead{$\lambda$8662} & \colhead{$\lambda$11690} & \colhead{$\lambda$11770} & \colhead{$\lambda$11773} & \colhead{$\lambda$11783} & \colhead{$\lambda$11828} & \colhead{$\lambda$11883} & \colhead{$\lambda$11893} & \colhead{$\lambda$11973} & \colhead{$\lambda$12432} & \colhead{$\lambda$12522} & \colhead{$\lambda$12679} & \colhead{$\lambda$12738} & \colhead{$\lambda$12822} & \colhead{$\lambda$12847} & \colhead{$\lambda$12900} & \colhead{$\lambda$13123} & \colhead{$\lambda$13135} & \colhead{$\lambda$13151}
}
\startdata
NGC 1316 & 0.00601 & 206.0 & -25.39 & 1.19 & 3.27 & 2.48 & 0.37 & 0.00 & 0.46 & 0.35 & 1.02 & 0.69 & 0.32 & 1.62 & 1.09 & 0.37 & 0.94 & 0.11 & 1.21 & 0.20 & 0.95 & 1.26 & 0.00 & 0.98\\
       & (0.00001) & (12.7) & (0.02) & (0.03) & (0.12) & (0.10) & (0.02) & (0.00) & (0.04) & (0.03) & (0.03) & (0.04) & (0.02) & (0.08) & (0.05) & (0.01) & (0.04) & (0.01) & (0.05) & (0.02) & (0.05) & (0.07) & (0.05) & (0.04)\\
NGC 1332 & 0.00540 & 353.0 & -24.00 & 0.73 & 2.73 & 1.71 & 0.00 & 1.02 & 0.18 & 0.00 & 0.74 & 0.35 & 0.00 & 1.46 & 1.33 & 0.11 & 1.26 & 0.00 & 0.95 & 0.33 & 0.99 & 2.22 & 0.00 & 0.98\\
       & (0.00002) & (45.2) & (0.02) & (0.03) & (0.21) & (0.12) & (0.00) & (0.23) & (0.14) & (0.04) & (0.06) & (0.05) & (0.00) & (0.11) & (0.11) & (0.02) & (0.12) & (0.01) & (0.18) & (0.12) & (0.09) & (0.20) & (0.14) & (0.10)\\
NGC 3258 & 0.00960 & 319.0 & -23.79 & 0.88 & 1.99 & 3.47 & 0.09 & 0.07 & 0.10 & 0.00 & 0.73 & 0.17 & 0.00 & 0.58 & 1.33 & 0.81 & 1.07 & 1.15 & 0.82 & 0.10 & 1.17 & 1.84 & 0.00 & 0.64\\
       & (0.00003) & (40.8) & (0.02) & (0.08) & (0.16) & (0.22) & (0.05) & (0.17) & (0.15) & (0.00) & (0.09) & (0.05) & (0.00) & (0.05) & (0.16) & (0.07) & (0.09) & (0.11) & (0.10) & (0.06) & (0.12) & (0.16) & (0.07) & (0.09)\\
NGC 3557 & 0.01027 & 233.0 & -25.27 & 0.30 & 1.77 & 3.33 & 0.24 & 0.00 & 0.49 & 0.00 & 0.72 & 0.13 & 0.00 & 1.21 & 0.85 & 0.16 & 0.76 & 0.78 & 1.09 & 0.58 & 0.67 & 2.11 & 0.23 & 1.09\\
       & (0.00002) & (16.2) & (0.02) & (0.01) & (0.06) & (0.15) & (0.03) & (0.02) & (0.06) & (0.01) & (0.05) & (0.04) & (0.00) & (0.08) & (0.05) & (0.02) & (0.02) & (0.03) & (0.04) & (0.01) & (0.04) & (0.15) & (0.08) & (0.04)\\
NGC 5845 & 0.00555 & 239.0 & -21.75 & 1.70 & 2.58 & 1.49 & 0.86 & 0.66 & 0.03 & 0.23 & 0.97 & 0.53 & 0.00 & 1.10 & 0.91 & 0.00 & 0.61 & 0.02 & 1.25 & 0.00 & 0.97 & 1.10 & 0.27 & 0.21\\
       & (0.00001) & (9.4) & (0.02) & (0.02) & (0.06) & (0.02) & (0.02) & (0.02) & (0.05) & (0.04) & (0.02) & (0.07) & (0.00) & (0.02) & (0.06) & (0.00) & (0.01) & (0.01) & (0.04) & (0.00) & (0.05) & (0.04) & (0.00) & (0.02)\\
NGC 7014 & 0.01637 & 305.0 & -24.34 & 0.51 & 2.66 & 1.18 & 0.54 & 0.00 & 0.36 & 0.20 & 0.55 & 0.84 & 0.16 & 1.26 & 1.41 & 0.26 & 2.11 & 1.34 & 1.61 & 0.00 & 2.10 & 1.07 & 0.62 & 0.00\\
       & (0.00003) & (35.0) & (0.03) & (0.06) & (0.27) & (0.12) & (0.13) & (0.11) & (0.17) & (0.16) & (0.13) & (0.14) & (0.16) & (0.15) & (0.16) & (0.17) & (0.21) & (0.16) & (0.19) & (0.17) & (0.12) & (0.24) & (0.26) & (0.19)\\
NGC 7410 & 0.00597 & 124.0 & -24.04 & 1.40 & 3.03 & 1.96 & 0.38 & 0.42 & 0.09 & 0.26 & 1.04 & 0.10 & 0.29 & 0.85 & 0.75 & 0.53 & 0.60 & 0.27 & 0.00 & 0.14 & 0.76 & 1.07 & 0.17 & 1.06\\
       & (0.00001) & (5.7) & (0.02) & (0.03) & (0.10) & (0.05) & (0.02) & (0.01) & (0.01) & (0.01) & (0.03) & (0.02) & (0.01) & (0.03) & (0.03) & (0.02) & (0.03) & (0.02) & (0.00) & (0.01) & (0.03) & (0.04) & (0.03) & (0.03)\\
NGC 7743 & 0.00566 &  85.7 & -22.74 & 0.98 & 1.99 & 1.76 & 0.03 & 0.36 & 0.00 & 0.48 & 0.80 & 0.44 & 0.37 & 0.96 & 0.57 & 0.27 & 1.08 & 0.22 & 0.98 & 0.08 & 0.74 & 0.95 & 0.09 & 1.11\\
       & (0.00001) & (8.2) & (0.03) & (0.05) & (0.14) & (0.12) & (0.01) & (0.01) & (0.01) & (0.02) & (0.04) & (0.02) & (0.01) & (0.06) & (0.07) & (0.02) & (0.09) & (0.02) & (0.11) & (0.02) & (0.04) & (0.05) & (0.03) & (0.05)\\
\end{deluxetable*}
\end{turnpage}

\begin{deluxetable}{rcccccc}
\tablecaption{EW Trend Fit Results\label{tbl:TrendParams}}
\tablehead{
\colhead{Line} & \colhead{$m_{\rm Mag}$} & \colhead{$b_{\rm Mag}$} & \colhead{$\chi^2_{\rm Mag}$} &\colhead{$m_{\sigma}$} & \colhead{$b_{\sigma}$} &\colhead{$\chi^2_{\sigma}$}
}
\startdata
Ca II      & -0.0485  &  0.78  & 6.75 & -0.0006  &  2.06  & 7.06 \\
           & (0.0713) & (1.69) &      & (0.0013) & (0.29) &      \\
K I        & -0.0040  &  0.32  & 1.97 &  0.0008  &  0.25  & 0.95 \\
           & (0.0291) & (0.70) &      & (0.0003) & (0.07) &      \\
Fe I       & -0.0098  &  0.30  & 7.51 & -0.0007  &  0.68  & 7.00 \\
           & (0.0674) & (1.61) &      & (0.0008) & (0.18) &      \\
Mg I       &  0.0078  &  1.08  & 2.63 & -0.0007  &  1.04  & 2.20 \\
           & (0.0404) & (0.96) &      & (0.0006) & (0.14) &      \\
Ti I       & -0.0456  & -0.83  & 2.94 &  0.0008  &  0.13  & 2.90 \\
           & (0.0451) & (1.09) &      & (0.0006) & (0.12) &      \\
Na I       & -0.0551  & -0.53  & 7.32 &  0.0018  &  0.42  & 6.89 \\
           & (0.0609) & (1.45) &      & (0.0014) & (0.31) &      \\
Pa $\beta$ &  0.0158  &  1.50  & 2.45 & -0.0001  &  1.14  & 2.50 \\
           & (0.0478) & (1.15) &      & (0.0012) & (0.29) &      \\
Mn I       &  0.0197  &  1.33  & 5.43 &  0.0018  &  0.51  & 3.62 \\
           & (0.0708) & (1.69) &      & (0.0010) & (0.20) &      \\
Al I       & -0.1580  & -2.69  & 5.66 &  0.0007  &  0.89  & 11.14\\
           & (0.0645) & (1.52) &      & (0.0015) & (0.31) &      \\
Ca I       &  0.0389  &  1.08  & 1.25 &  0.0003  &  0.12  & 1.46 \\
           & (0.0325) & (0.74) &      & (0.0005) & (0.11) &      \\
\end{deluxetable}

\begin{turnpage}
  \tabletypesize{\footnotesize}
  \setlength{\tabcolsep}{0.04in}
  \begin{deluxetable*}{ccccccccccccccccccccccccccc}
    \tabletypesize{\small}

    \tablecaption{Model-Predicted Equivalent Widths \AA}
    \label{tbl:ModLines}
    \vspace*{5 mm}
    \tablehead{\colhead{IMF}\tablenotemark{a}& 
      \colhead{Ca II}&
      \colhead{Ca II}&
      \colhead{Ca II}&
      \colhead{K I}&
      \colhead{K I}&
      \colhead{K I}&
      \colhead{Fe I}&
      \colhead{Mg I}&
      \colhead{Fe I}&
      \colhead{Ti I}&
      \colhead{Fe I}&
      \colhead{K I}&
      \colhead{K I}&
      \colhead{Na I}&
      \colhead{Ti I}&
      \colhead{Pa $\beta$}&
      \colhead{Ti I}&
      \colhead{Mn I}&
      \colhead{Al I}&
      \colhead{Ca I}&
      \colhead{Al I}&
      \colhead{$\sum$(Ca II)} &
      \colhead{$\sum$(K I)} &
      \colhead{$\sum$(Fe I)} &
      \colhead{$\sum$(Ti I)} &
      \colhead{$\sum$(Al I)}}

    \startdata
    \sidehead{$Z = Z_\odot$, Age = 10 Gyr}
    0  &0.56 &1.27 &1.20 &0.52 &0.04 &0.77 &0.12 &0.69 &1.17 &0.36 &0.92 &0.25 &0.17 &0.19 &0.21 &0.48 &0.12 &0.16 &0.94 &0.10 &0.48 &3.03 &1.75 &2.22 &0.69 &1.43\\
    1  &0.56 &1.29 &1.22 &0.45 &0.03 &0.64 &0.09 &0.66 &1.14 &0.34 &0.91 &0.21 &0.13 &0.18 &0.20 &0.51 &0.10 &0.15 &0.92 &0.09 &0.42 &3.07 &1.46 &2.15 &0.64 &1.35\\
    2  &0.61 &1.39 &1.30 &0.49 &0.03 &0.67 &0.09 &0.70 &1.20 &0.34 &0.97 &0.21 &0.16 &0.18 &0.20 &0.52 &0.10 &0.15 &1.02 &0.09 &0.46 &3.31 &1.56 &2.27 &0.64 &1.48\\
    3  &0.56 &1.29 &1.22 &0.45 &0.03 &0.61 &0.09 &0.61 &1.06 &0.31 &0.84 &0.21 &0.13 &0.15 &0.17 &0.47 &0.09 &0.12 &0.83 &0.09 &0.40 &3.07 &1.43 &1.99 &0.57 &1.23\\
    4  &0.61 &1.39 &1.30 &0.49 &0.03 &0.67 &0.09 &0.70 &1.20 &0.34 &0.97 &0.21 &0.16 &0.18 &0.20 &0.52 &0.10 &0.15 &1.02 &0.09 &0.46 &3.31 &1.56 &2.27 &0.64 &1.48\\
    5  &0.60 &1.18 &1.17 &0.52 &0.17 &0.80 &0.27 &0.74 &1.19 &0.40 &0.89 &0.33 &0.32 &0.31 &0.21 &0.40 &0.15 &0.26 &1.14 &0.17 &0.78 &2.94 &2.15 &2.35 &0.75 &1.92\\
    \sidehead{$Z = Z_\odot/3$, Age = 10 Gyr}
    0  &0.56 &1.27 &1.21 &0.51 &0.07 &0.79 &0.12 &0.68 &1.14 &0.37 &0.91 &0.24 &0.17 &0.16 &0.22 &0.48 &0.10 &0.15 &1.04 &0.09 &0.50 &3.04 &1.79 &2.16 &0.68 &1.54\\
    1  &0.56 &1.30 &1.23 &0.44 &0.05 &0.64 &0.09 &0.64 &1.10 &0.33 &0.89 &0.20 &0.13 &0.16 &0.20 &0.49 &0.09 &0.14 &0.98 &0.09 &0.43 &3.09 &1.45 &2.08 &0.62 &1.40\\
    2  &0.60 &1.37 &1.31 &0.44 &0.05 &0.67 &0.10 &0.66 &1.16 &0.37 &0.95 &0.20 &0.13 &0.16 &0.23 &0.52 &0.10 &0.17 &1.10 &0.13 &0.47 &3.27 &1.49 &2.21 &0.70 &1.57\\
    3  &0.58 &1.28 &1.23 &0.40 &0.05 &0.57 &0.08 &0.57 &1.00 &0.30 &0.82 &0.17 &0.11 &0.13 &0.17 &0.46 &0.09 &0.12 &0.92 &0.09 &0.41 &3.09 &1.29 &1.91 &0.56 &1.32\\
    4  &0.60 &1.37 &1.31 &0.44 &0.05 &0.67 &0.10 &0.66 &1.16 &0.37 &0.95 &0.20 &0.13 &0.16 &0.23 &0.52 &0.10 &0.17 &1.10 &0.13 &0.47 &3.27 &1.49 &2.21 &0.70 &1.57\\
    5  &0.60 &1.19 &1.18 &0.54 &0.17 &0.84 &0.24 &0.73 &1.16 &0.40 &0.88 &0.33 &0.32 &0.27 &0.21 &0.39 &0.13 &0.23 &1.15 &0.14 &0.73 &2.97 &2.20 &2.28 &0.74 &1.88\\
    \sidehead{$Z = Z_\odot$, Age = 13 Gyr}
    0  &0.53 &1.22 &1.14 &0.47 &0.04 &0.79 &0.16 &0.67 &1.12 &0.34 &0.89 &0.22 &0.15 &0.17 &0.22 &0.40 &0.12 &0.17 &1.00 &0.09 &0.57 &2.89 &1.66 &2.17 &0.67 &1.57\\
    1  &0.52 &1.23 &1.16 &0.41 &0.03 &0.65 &0.14 &0.63 &1.08 &0.32 &0.88 &0.18 &0.12 &0.16 &0.21 &0.42 &0.10 &0.16 &0.98 &0.09 &0.51 &2.91 &1.40 &2.10 &0.62 &1.49\\
    2  &0.58 &1.34 &1.22 &0.41 &0.03 &0.68 &0.14 &0.65 &1.12 &0.32 &0.91 &0.18 &0.12 &0.18 &0.23 &0.43 &0.10 &0.16 &1.07 &0.09 &0.55 &3.14 &1.43 &2.17 &0.64 &1.63\\
    3  &0.51 &1.22 &1.09 &0.38 &0.03 &0.60 &0.13 &0.58 &1.01 &0.30 &0.82 &0.16 &0.12 &0.16 &0.18 &0.40 &0.09 &0.16 &0.93 &0.09 &0.49 &2.82 &1.29 &1.96 &0.57 &1.42\\
    4  &0.58 &1.34 &1.22 &0.41 &0.03 &0.68 &0.14 &0.65 &1.12 &0.32 &0.91 &0.18 &0.12 &0.18 &0.23 &0.43 &0.10 &0.16 &1.07 &0.09 &0.55 &3.14 &1.43 &2.17 &0.64 &1.63\\
    5  &0.57 &1.15 &1.12 &0.46 &0.12 &0.80 &0.24 &0.72 &1.14 &0.39 &0.86 &0.27 &0.26 &0.26 &0.21 &0.34 &0.14 &0.22 &1.09 &0.13 &0.75 &2.84 &1.91 &2.24 &0.74 &1.84\\
    \enddata
    \tablenotetext{a}{IMF codes are as follows: 0 -- \citet{sal55}, 1 -- \citet{cha03}, 2 -- \citet{kro01}, 3 -- \citet{van08}, 4 -- \citet{dav08}, 5 -- Dwarf-rich variant}
  \end{deluxetable*}

\end{turnpage}

\end{document}